\documentclass[a4paper,twocolumn,accepted=2025-04-03]{quantumarticle}
\pdfoutput=1
\usepackage[utf8]{inputenc}
\usepackage[english]{babel}
\usepackage[table]{xcolor}
\usepackage[T1]{fontenc}
\usepackage[numbers,sort&compress]{natbib}
\usepackage[colorlinks=true]{hyperref}
\usepackage{tabularx}
\usepackage{multirow}
\usepackage{booktabs}
\usepackage{graphicx}
\usepackage{amsmath}
\usepackage{amssymb}
\usepackage{amsthm}
\usepackage{mathtools}
\usepackage{bm}
\usepackage{quantikz}
\usepackage{tikz}
\usepackage{caption}
\usepackage{subcaption}
\usepackage{physics}
\usepackage{siunitx}
\usepackage{xspace}
\usepackage{soul}

\newcommand{\im}{\mathrm{i}}
\newcommand{\e}{\mathrm{e}}

\newcommand{\diff}{\mathrm{d}}
\newcommand{\CnZ}[1]{C$^{#1} Z$}
\newcommand{\CbZ}{\CnZ{b}\xspace}
\newcommand{\CX}{C$X$\xspace}
\newcommand{\Rz}{\ensuremath{R_Z}\xspace}
\newcommand{\Rx}{\ensuremath{R_X}\xspace}
\newcommand{\RZZ}{\ensuremath{ZZ\text{Phase}}\xspace}
\newcommand{\PhasedX}{\ensuremath{\text{Phased}X}\xspace}
\DeclareMathOperator{\poly}{poly}

\definecolor{NormalBlue}{HTML}{4053d3}
\definecolor{NormalRed}{HTML}{b51d14}
\definecolor{NormalYellow}{HTML}{ddb310}
\definecolor{NormalCyan}{HTML}{00beff}

\newcommand{\eg}{e.g.\xspace}

\begin{document}

\title{Quantum state preparation for multivariate functions}

\author[1]{Matthias Rosenkranz}
\orcid{0000-0002-1605-9141}
\email{matthias.rosenkranz@quantinuum.com}

\author[1]{Eric Brunner}
\orcid{0000-0001-7631-6528}

\author[1]{Gabriel Marin-Sanchez}
\orcid{0009-0003-7630-6081}

\author[2]{Nathan Fitzpatrick}
\orcid{0000-0001-5819-9129}

\author[2]{Silas Dilkes}
\orcid{0000-0003-2186-0379}

\author[2]{Yao Tang}
\orcid{0009-0006-4023-8701}

\author[3,4]{Yuta Kikuchi}
\orcid{0000-0002-1802-5260}

\author[1]{Marcello Benedetti}
\orcid{0000-0003-0231-1729}
\email{marcello.benedetti@quantinuum.com}

\affil[1]{Quantinuum, Partnership House, Carlisle Place, London SW1P 1BX, United Kingdom}

\affil[2]{Quantinuum, Terrington House, 13-15 Hills Road, Cambridge CB2 1NL, United Kingdom}

\affil[3]{Quantinuum K.K., Otemachi Financial City Grand Cube 3F, 1-9-2 Otemachi, Chiyoda-ku, Tokyo, Japan}

\affil[4]{Interdisciplinary Theoretical and Mathematical Sciences Program (iTHEMS), RIKEN, Wako, Saitama 351-0198, Japan}

\maketitle

\begin{abstract}
A fundamental step of any quantum algorithm is the preparation of qubit registers in a suitable initial state. Often qubit registers represent a discretization of continuous variables and the initial state is defined by a multivariate function. We develop protocols for preparing quantum states whose amplitudes encode multivariate functions by linearly combining block-encodings of Fourier and Chebyshev basis functions. Without relying on arithmetic circuits, quantum Fourier transforms, or multivariate quantum signal processing, our algorithms are simpler and more effective than previous proposals. We analyze requirements both asymptotically and pragmatically in terms of near/medium-term resources. Numerically, we prepare bivariate Student's t-distributions, 2D Ricker wavelets and electron wavefunctions in a 3D Coulomb potential, which are initial states with potential applications in finance, physics and chemistry simulations. Finally, we prepare bivariate Gaussian distributions on the Quantinuum H2-1 trapped-ion quantum processor using 24 qubits and up to 237 two-qubit gates.
\end{abstract}

\section{Introduction}

State preparation is a computational problem that appears as a sub-routine in many quantum algorithms. In Hamiltonian simulation, we approximate the real-time evolution $\e^{-\im t H} \ket{\psi}$ of an initially prepared state $\ket{\psi}$. Quantum algorithms for linear systems of equations $A \ket{x} = \ket{b}$ require the preparation of $\ket{b}$ in the first place. Other examples are found in the literature of quantum algorithms for partial differential equations~\cite{leyton2008quantum,Childs2021highprecision,costaQuantumAlgorithmSimulating2019}, financial instrument pricing~\cite{herman2022survey,stamatopoulos2023derivative,akhalwaya2023modular}, quantum chemistry simulations~\cite{AspuruGuzik_2005,Kassal_2008,Cao_2019,McArdle_2020,Liu_2022,grid1q}, high-energy or nuclear physics~\cite{Jordan_2012,Banuls_2020,Bauer_2023}, and other applications~\cite{Dalzell_2023}.

The computational complexity of state preparation is well understood~\cite{Vartiainen_2004, Mottonen_2004, Bergholm_2005, Shende_2006, Plesch_2011, low2018trading} and it is connected to that of synthesizing circuits for arbitrary unitary matrices. Algorithms for preparing generic $n$-qubit states incur an unavoidable exponential scaling, either in terms of circuit depth~\cite{Sun_2023} or number of ancilla qubits~\cite{Zhang_2022}. Under certain assumptions, however, the scaling can be improved. For the special case of $k$-sparse states, i.e. states with $k$ non-zero amplitudes, there exists an efficient preparation using $\log(nk)$ circuit depth and $\mathcal{O}(nk \log k)$ ancilla qubits~\cite{Zhang_2022}. For more structured states, such as when the amplitudes are defined by a function, there exist a number of efficient algorithms. For example, \cite{zalkaSimulatingQuantumSystems1998,grover2002creating, kitaev2009wavefunction, bauer2021practical, Rattew2021efficient, Marin-Sanchez_2023} consider certain classes of probability density functions; \cite{Holmes_2020, Gonzalezconde_2023, Iaconis_2024} consider real-valued functions approximated by polynomials; \cite{Rattew_2022} considers complex-valued continuous functions. Quantum arithmetic circuits are among the most expensive components in state preparation algorithms. Recent work has aimed at limiting their use, potentially leading to a reduction of the number of qubits and gates. When the amplitudes are unknown and provided by a black-box, the use of arithmetic circuits can be limited to simple comparisons against some initial reference state~\cite{Sanders_2019,Bausch2022fastblackboxquantum,lemieux2024quantum}, or replaced by linear combination of unitaries circuits~\cite{Wang_2021}. When amplitudes are given by a known function instead, quantum signal processing (QSP)~\cite{Low_2017, Gilyen_2019, Silva2021, Martyn2021, Kikuchi_2023} can be used to avoid arithmetic circuits altogether~\cite{mcardle2022quantum}.

In this work, we contribute to the state preparation problem by considering the case where the amplitudes are defined by multivariate functions. This is an essential step for many quantum computing applications of practical interest. For example, quantum algorithms for simulating wave propagation in $D$-dimensional space requires preparation of a $D$-dimensional function in the amplitudes of a quantum state as initial condition~\cite{costaQuantumAlgorithmSimulating2019}. Another example is the simulation of quantum chemistry in first quantization~\cite{grid1q}. Here, a typical initial state is an antisymmetrized many-body wave function represented in a $D$-dimensional discrete Fourier basis, which can be understood as a multivariate function. Despite its practical importance, the multivariate state preparation problem is rarely discussed in the literature and presents its unique challenges. For example, multivariate versions of QSP appear to be much more complicated and highly constrained~\cite{Rossi2022, mori2023comment, nemeth2023variants}. We attack the multivariate state preparation problem using a simple set of primitives and without using arithmetic circuits. We make use of (i) function approximation by truncated Fourier or Chebyshev series, (ii) block-encoding of the corresponding basis functions, and (iii) linear combination of unitaries. The resources required by our algorithms depend on three hyper-parameters: the number of dimensions $D$, the number of bits $n$ used for the discretization of each dimension, and the degree $d$ used for function approximation in each dimension. We obtain a number of two-qubit gates that scales as $\mathcal{O}(d^D + Dn \log d)$ for Fourier series and as $\mathcal{O}\left(d^D + Ddn\log n\right)$ for Chebyshev series. The total number of qubits scales as $\mathcal{O}\left(Dn + D\log d\right)$ for both methods.

Some of the primitives we use were originally introduced by Childs et al.~\cite{Childs_2017} in the context of quantum linear system solvers. In the context of state preparation the closest work to ours is the Fourier series loader (FSL)~\cite{Moosa2023}. This method loads the Fourier coefficients in the main qubit register, hence avoiding the use of ancilla qubits. It then employs the quantum Fourier transform (QFT) to prepare the target state using additional $\mathcal{O}(n^2)$ gates. Related ideas of using the QFT for Fourier series interpolation of a target function from a coarse grid to a finer grid were earlier explored in~\cite{garcia-ripollQuantuminspiredAlgorithmsMultivariate2021}. In comparison, our method uses ancilla qubits, but it is more general, and, in the limit of large $n$, cuts down the number of two-qubit gates by avoiding QFT. A related method called the Walsh series loader (WSL)~\cite{zylberman2023efficient} approximates the target function by a Walsh series, which is loaded on the quantum computer using techniques from~\cite{Welch_2014}. While WSL is shallower than FSL, it introduces a further approximation error which also affects the probability of success. Our method does not have this additional error. We provide a thorough comparison of multivariate state preparation techniques  in Appendix~\ref{app:prior_works}. The code implementing the presented protocols and the data to reproduce all results are available at GitHub and Zenodo~\cite{mvsp1.0.0.2025}.

\section{Methods}\label{sec:methods}

We wish to prepare a quantum state with amplitudes proportional to a multivariate target function. We consider complex target functions, $f\colon [-1, 1]^D \rightarrow \mathbb{C}$. We assume that $f$ admits a Fourier or Chebyshev series representation. A Lipshitz condition on $f$ is sufficient for uniform convergence of this series (in the Fourier case we also assume periodicity on $[-1,1]^D$)~\cite{masonNearbestMultivariateApproximation1980}. For simplicity of exposition, let us first discuss the one-dimensional case, $D=1$.  The Fourier and Chebyshev series of $f$ are $f(x) = \sum_{k=-\infty}^\infty \hat{c}_{k} \e^{\im\pi k x}$ and $f(x) = \sum_{k=0}^\infty \hat{c}_{k} T_{k}(x)$, respectively, with $T_k(x)$ the $k$-th Chebyshev polynomial of the first kind. For the Fourier case, this requires the function to be periodic with period $2$ on the whole interval $[-1,1]$. However, we also work with non-periodic target functions as Fourier series by focusing on the interval $[0, 1]$ and using the interval $[-1,0)$ for its periodic extension (see Appendix~\ref{app:preprocessing}).

Next, we approximate the target function with a finite Fourier or Chebyshev series of the form
\begin{alignat}{2}
    f_d^F(x) &=\sum_{k=-d}^{d} c_k \e^{\im\pi k x} &&\quad \text{ (Fourier)} \label{eq:f_d_Fourier}\\
    \intertext{and}
    f_d^C(x) &=\sum_{k=0}^d c_k T_k(x) &&\quad \text{ (Chebyshev)} \label{eq:f_d_Chebyshev},
\end{alignat}
respectively. The largest index in the sum, $d$, is the \emph{degree} of the polynomial (for which $c_d$ or $c_{-d}$ does not vanish). We use $f_d$ to indicate either $f_d^F$ or $f_d^C$, which should be clear from context. Two common methods for finding the coefficients $c_k$ are \emph{truncation} and \emph{interpolation}. If the coefficients of the infinite series are known, truncation sets $c_k=\hat c_k$ for all $|k|\leq d$.  Interpolation matches $f_d(x_j)$ with $f(x_j)$ in $d+1$ (Chebyshev) or $2d+1$ (Fourier) interpolation points $x_j$. Computing the exact coefficients for this interpolant can be done with the fast Fourier transform (FFT) with classical preprocessing cost $\mathcal{O}(d\log d)$ (see Appendix~\ref{app:Chebyshev}--\ref{app:preprocessing}). Interpolation is often advantageous because the series coefficients $\hat{c}_k$ may not be known analytically for target functions of practical interest and numerically integrating their definition, Eq.~\eqref{eq:cheb_coeff}, could introduce higher preprocessing cost or an additional error. The uniform error of interpolation is at most a factor of 2 larger than the uniform error of truncation~\cite{boydChebyshevFourierSpectral2001}. A rule-of-thumb is that the smoother the target function the faster its finite series approximations converges with increasing $d$~\cite{boydChebyshevFourierSpectral2001,trefethenApproximationTheoryApproximation2019,trefethenMultivariatePolynomialApproximation2017}.

For the multivariate case, we consider approximations with products of the corresponding basis functions. For example, in two dimensions
\begin{align}
    f_{d_x,d_y}^F(x, y) &=\sum_{k=-d_x}^{d_x} \sum_{l=-d_y}^{d_y} c_{k,l} \e^{\im\pi k x} \e^{\im\pi l y} \quad\text{(Fourier)} \label{eq:f_d_Fourier_2d}\\
    \intertext{and}
    f_{d_x,d_y}^C(x, y) &=\sum_{k=0}^{d_x} \sum_{l=0}^{d_y} c_{k,l} T_k(x) T_l(y) \quad\text{(Chebyshev)}.\label{eq:f_d_Chebyshev_2d}
\end{align}
This generalizes to $D$ dimensions with degrees $\bm{d}=(d_1, d_2, \dots, d_D)$. The largest value in $\bm{d}$ is the \emph{maximal degree} of the polynomial. The following protocols prepare a quantum state $\sum_{\bm{x}} g_{\bm{d}}(\bm{x}) \ket{\bm{x}}$ with $g_{\bm{d}}(\bm{x}) = f_{\bm{d}}(\bm{x}) / \sqrt{\sum_{\bm{x}} |f_{\bm{d}}(\bm{x})|^2}$ for all $\bm{x}$ on a grid of size $2^{n_1} \times 2^{n_2} \times \cdots \times 2^{n_D}$. A generic way to compute coefficients of $f_{\bm{d}}$ is via the multi-dimensional FFT with preprocessing cost $\mathcal{O}[\bar{d} \log(\bar{d})]$ requiring at most $\bar{d}$ function evaluations, where $\bar d = \Pi_{j=1}^D d_j$. For more structured multivariate functions, other ways of finding coefficients may be more efficient in practice. For example, one could attempt to compute a low-rank approximation of $f$, e.g. via iterative Gaussian elimination, adaptive cross approximation etc. (see e.g. Ref.~\cite{townsendExtensionChebfunTwo2013} for their usage in classical numerical computation). Our protocols are agnostic to this preprocessing and only require the coefficients of a finite Fourier or Chebyshev series as inputs.

\subsection{Discretization}
\label{sec:discretization}

In this work, each variable is assigned to its own `main' register. We use the notation $x$ for the variable and $n_x$ for the number of qubits in the register. The variable is therefore discretized on a grid of $2^{n_x}$ points. In the univariate setting, we simplify the notation to $n = n_x$.

Recall that when working with Fourier series approximations, we focus on the non-negative part of the domain, $x \in [0,1]$. 
In this case the uniform grid of points is represented by the $n$-qubit diagonal matrix
\begin{equation}
\label{eq:grid_fourier}
    H^{\mathrm{F}} = \frac{1}{2^{n}-1} \mqty(\dmat{0,1,\ddots,2^n -1}).
\end{equation}
For example, with $n=2$ we have grid points $0, \tfrac{1}{3}, \tfrac{2}{3}, 1$. 

When working with Chebyshev series approximations we don't need periodic extensions. Thus we use the negative part of the domain as well, $x \in [-1,1]$. The uniform grid of point is then represented by the $n$-qubit diagonal matrix
\begin{equation}
\label{eq:grid_chebyshev}
    H^{\mathrm{C}} = 2 H^{\mathrm{F}} - I^{\otimes{n}} ,
\end{equation}
where $I = \mqty(\pmat{0})$ is the identity matrix. For example, with $n=2$ we have grid points $-1, -\tfrac{1}{3}, \tfrac{1}{3}, 1$. It is important to note that this grid is uniform and should not be confused with the grid of Chebyshev nodes commonly used in polynomial interpolation. 

In the following we often use $H$ to indicate either $H^{\mathrm{F}}$ or $H^{\mathrm{C}}$, which should be clear from the context. Note that both matrices \eqref{eq:grid_fourier} and \eqref{eq:grid_chebyshev} have the operator norm $\| H \| = 1$. Grids for multivariate functions are built by taking tensor products $H \otimes \cdots \otimes H$, and this does not change the operator norm. 

\subsection{Block-encoding}
\label{sec:block_encodings}

To construct the Fourier and Chebyshev approximations we shall block-encode the corresponding basis functions. 
We begin with the standard definition of block-encoding.
An $(n+m)$-qubit unitary $W_L$ is an $(\lambda, m, \epsilon)$-block-encoding of $L \in \mathbb{C}^{2^n \times 2^n}$ if
\begin{equation}
\label{eq:block_encoding}
    \left \| L - \lambda \left( \bra{0}^{\otimes m} \otimes I^{\otimes n} \right) W_L \left( \ket{0}^{\otimes m} \otimes I^{\otimes n} \right) \right \| \leq \epsilon,
\end{equation}
where $m$ is the number of ancilla qubits, $\lambda$ is a normalization constant such that $\lambda \geq \Vert L \Vert - \epsilon$  and $\epsilon$ is the error.

We make use of the linear combination of unitaries (LCU)~\cite{Childs2012} to construct block-encoding circuits. Assume the input matrix is given as a weighted sum of $K$ unitary operators~$\{U_k\}_k$, i.e. $L = \sum_{k=0}^{K-1} \lambda_k U_k$. To deal with negative and complex coefficients, we write them in polar form, $\lambda_k = \e^{\im \gamma_k} |\lambda_k|$, and compute the normalization $\lambda = \sum_{k=0}^{K-1} |\lambda_k|$. LCU uses $m = \lceil \log_2 K \rceil$ ancilla qubits and the following unitary operators
\begin{align}
    A &= \sum_{k=0}^{K-1} \sqrt{\frac{|\lambda_k|}{\lambda}} \dyad{k}{0}^{\otimes m} \otimes I^{\otimes n} + \text{ u.c. },\label{eq:lcu_operator_A}\\
    B &= \sum_{k=0}^{K-1} \dyad{k} \otimes U_k,\label{eq:lcu_operator_B}\\
    C &= \sum_{k=0}^{K-1} \e^{\im \gamma_k} \dyad{k} \otimes I^{\otimes n},\label{eq:lcu_operator_C}
\end{align}
where u.c. stands for unitary completion. With these definitions, the operator $A^\dag C B A$ yields
\begin{multline}
\label{eq:lcu}
    \left( \bra{0}^{\otimes m} \otimes I^{\otimes n} \right) \; A^\dag C B A \; \left( \ket{0}^{\otimes m} \otimes I^{\otimes n} \right) =\\
    \frac{1}{\lambda} \sum_{k=0}^{K-1} \lambda_k U_k ,
\end{multline}
which is a $(\lambda, m, 0)$-block-encoding of $L$.

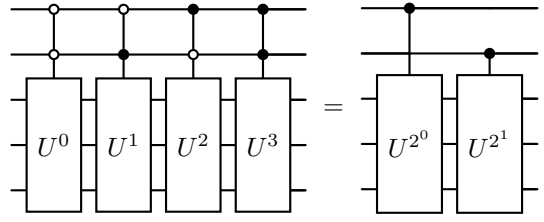
\begin{figure}[t]
\centering
\begin{quantikz}[row sep={0.6cm, between origins}, column sep=0.2cm]
     & \octrl{1} & \octrl{1} & \ctrl{1} & \ctrl{1} & \qw\\
     & \octrl{1} & \ctrl{1} & \octrl{1} & \ctrl{1} & \qw \\
     & \gate[3]{U^0} & \gate[3]{U^1} & \gate[3]{U^2} & \gate[3]{U^3} & \qw\\
     & & & & & \qw\\
     & & & & & \qw
\end{quantikz}
= 
\begin{quantikz}[row sep={0.6cm, between origins}, column sep=0.2cm]
         & \ctrl{2} & \qw & \qw\\
         & \qw & \ctrl{1} &  \qw\\
         & \gate[3]{U^{2^0}} & \gate[3]{U^{2^1}} & \qw\\
         & & & \qw\\
         & & & \qw
\end{quantikz}
\caption{Circuit identity Eq.~\eqref{eq:uid} for $\sum_k \dyad{k} \otimes U^k$ with $K=4$ and $m=2$.}
\label{fig:identity}
\end{figure}

Let us discuss the important special case where $U_k = U^k$ are powers of some unitary $U$. When $k\geq 0$ is encoded to the state as a binary string $\ket{k}=\ket{k_{m-1}}\otimes \dots \otimes\ket{k_0}$, corresponding to the binary representation $k=2^0 k_0 + 2^1 k_1 +\cdots + 2^{m-1}k_{m-1}$, the $B$ operator in Eq.~\eqref{eq:lcu_operator_B} takes the form
\begin{equation}
\label{eq:uid}
   \sum_{k = 0}^{K-1} \dyad{k} \otimes U^k 
   = \prod_{i=0}^{m-1} \left[ \dyad{0}_i \otimes I^{\otimes n} + \dyad{1}_i \otimes U^{2^i} \right]
\end{equation}
where we highlight the relevant qubits with a subscript and assume identity gates on all other qubits. The identity in Eq.~\eqref{eq:uid} is also discussed in Lemma 8 of~\cite{Childs_2017}, and is implicitly used in many other quantum algorithms, e.g. it is a key component of quantum phase estimation.  This simple identity is illustrated in Fig.~\ref{fig:identity}. 

Note that the approximate functions we wish to encode, Eqs.~\eqref{eq:f_d_Fourier} and \eqref{eq:f_d_Chebyshev}, are weighted sums of basis functions. In the following sections we implement the $k$-th basis functions as the operator $U_k$ and then use LCU to block-encode the whole weighted sum. The circuit construction for the operators $A$ and $C$ does not change depending on whether the Fourier or Chebyshev basis is used, while the circuit construction for the operator $B$ does.

\begin{figure*}[tb]
\centering
\begin{tikzpicture} \node[scale=0.85] {
    \begin{quantikz}[row sep={0.7cm, between origins}, column sep=0.2cm]
         & \octrl{1} & \octrl{1} & \ctrl{1} & \ctrl{1} & \\
         & \octrl{1} & \ctrl{1} & \octrl{1} & \ctrl{1} & \\
         & \gate[3]{\e^{\im \pi 0 H^F}} & \gate[3]{\e^{\im \pi 1 H^F}} & \gate[3]{\e^{\im \pi 2 H^F}} & \gate[3]{\e^{\im \pi 3 H^F}} &\\
         & & & & &\\
         & & & & &
    \end{quantikz} 
    =
    \begin{quantikz}[row sep={0.7cm, between origins}, column sep=0.2cm]
         & \ctrl{2} & \ctrl{3} & \ctrl{4} & & & &  \\
         & & & & \ctrl{1} & \ctrl{2} & \ctrl{3} &  \\
         & \gate{P\left(\frac{\pi 2^{0+0}}{2^3 -1}\right)} & & & \gate{P\left(\frac{\pi 2^{0+1}}{2^3 -1}\right)} & & &  \\
         & & \gate{P\left(\frac{\pi 2^{1+0}}{2^3 -1}\right)} & & & \gate{P\left(\frac{\pi 2^{1+1}}{2^3 -1}\right)} & & \\
         & & & \gate{P\left(\frac{\pi 2^{2+0}}{2^3 -1}\right)} & & & \gate{P\left(\frac{\pi 2^{2+1}}{2^3 -1}\right)} &
    \end{quantikz}
};
\end{tikzpicture}
\caption{Example of circuit synthesis for the Fourier basis with Hamiltonian $H^F$, $a=2$ ancilla and $n=3$ main qubits. The special structure of this Hamiltonian leads to an efficient circuit made of $a \times n$ controlled-phase gates.}
\label{fig:factorize_fourier}
\end{figure*}
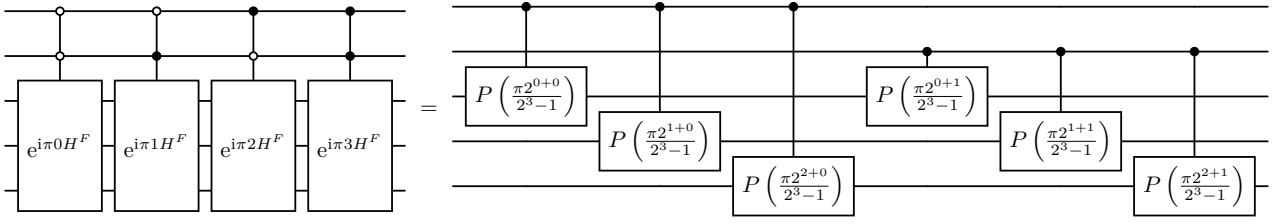
 
\subsection{Fourier basis functions} 
\label{sec:fourier}

We seek a circuit construction for the Fourier basis $B$ operator  $B^F=\sum_{k=0}^{K-1}\dyad{k}\otimes U_k$ with $U_k$ a block-encoding of the $k$-th Fourier basis function.
As the univariate Fourier basis in Eq.~\eqref{eq:f_d_Fourier} is defined as $\{ \e^{\im \pi k' x} \}_{k'=-d}^d$, the $k'$-th Fourier basis function applied to $H^F$ is the real-time evolution operator $U_{k'}=U^{k'}=\e^{\im \pi k' H^F}$. By the definition in Eq.~\eqref{eq:block_encoding}, a unitary operator is a trivial $(1, 0, 0)$-block-encoding of itself. However, we need to show that there exists an efficient circuit to implement the operator for any value of $k'$. It can be verified by inspection that 
\begin{align}
\label{eq:rte_operator}
    \e^{\im \pi k' H^F} = \bigotimes_{j=0}^{n-1} P_j \left( \frac{\pi k' 2^{j}}{2^{n} - 1} \right) ,
\end{align}
where 
\begin{align}
P(\theta) = \begin{pmatrix}
    1 & 0\\
    0 & \e^{\im \theta}
    \end{pmatrix} ,
\end{align}
is the phase shift gate and subscript $j$ indicates application of the gate to the $j$-th qubit. This is a block-encoding circuit for any Fourier basis function and uses only $n$ single-qubit phase shift gates. 

To map this block-encoding to the special case discussed at the end of Sec.~\ref{sec:block_encodings}, we first write $U_{k}=U^{-d} U^{k}$ with $k=0, 1, \dots, 2d$. Next we implement the operator $\sum_{k=0}^{K-1}\dyad{k} \otimes \e^{\im\pi kH^F}$ using Eq.~\eqref{eq:uid} with $U = \e^{\im \pi H^F}$ and an ancilla register of size $m = a = \log_2(2d+1)$, and synthesize the circuit efficiently as
\begin{align}
\label{eq:fourier_block_enc}
\begin{split}
   &\sum_{k=0}^{K-1}\dyad{k}\otimes \e^{\im \pi kH^F}=
   \\
   &\prod_{i=0}^{a-1} \prod_{j=0}^{n-1} \left[
  \dyad{0}_i \otimes I_j + \dyad{1}_i \otimes P_j \left(\frac{\pi2^{j+i}}{2^n - 1} \right) \right].
\end{split}
\end{align}
This can be verified using the factorization $\e^{\im \pi k H^F} = \prod_{i=0}^{a-1} \e^{\im \pi  k_i 2^i H^F}$, followed by Eq.~\eqref{eq:rte_operator}, and using $P_j(0) = I_j$. An example circuit diagram is provided in Fig.~\ref{fig:factorize_fourier}. Finally, we construct 
\begin{align}
\label{eq:fourier_b}
    B^F= U^{-d} \sum_{k=0}^{K-1}\dyad{k} \otimes \e^{\im\pi kH^F}
\end{align}
by applying the circuit Eq.~\eqref{eq:fourier_block_enc} followed by the circuit for $U^{-d}$ using Eq.~\eqref{eq:rte_operator} one more time. We have thus obtained a controlled $(1,0,0)$-block-encoding circuit for the Fourier basis functions. It requires only $a \times n$ controlled-phase gates and additional $n$ single-qubit gates from the circuit for $U^{-d}$. A controlled-phase gate can be implemented with two controlled-not (\CX)  gates plus single-qubit gates.

\subsection{Chebyshev basis functions}\label{sec:chebyshev}

We seek a circuit construction for the Chebyshev basis $B$ operator $B^C=\sum_{k=0}^{K-1}\dyad{k}\otimes U_k$ with $U_k$ a block-encoding of the $k$-th Chebyshev polynomial. As the univariate Chebyshev basis in Eq.~\eqref{eq:f_d_Chebyshev} is defined as $\{ T_k(x) = \cos(k \cos^{-1}(x)) \}_{k=0}^d$, the $k$-th Chebyshev basis function applied to $H^{\mathrm{C}}$ is not a unitary operator. We proceed by first block-encoding $H^{\mathrm{C}}$, and then using the algebra of block-encoding circuits to obtain $T_k(H^C)$.
 
Using Eqs.~\eqref{eq:grid_fourier} and \eqref{eq:grid_chebyshev}, we obtain 
\begin{align}
\label{eq:HC_block_encoding}
    H^C = \sum_{j=0}^{n-1} \left(- \frac{2^{j}}{2^n-1} \right) Z_j = \sum_{j=0}^{n-1} \left(\frac{2^{j}}{2^n-1} \right) X_jZ_jX_j.
\end{align}
Here $Z_j$ and $X_j$ indicate the Pauli operators $Z = \begin{pmatrix} 1 & 0\\ 0 & -1 \end{pmatrix}$ and $X  = \begin{pmatrix} 0 & 1\\ 1 & 0 \end{pmatrix}$ applied to the $j$-th qubit, respectively. The result is a linear combination of $n$ unitaries with which we can define a $(1, b, 0)$-block-encoding of $H^{\mathrm{C}}$,  $V = A^\dag_V C_V B_V A_V$, where $C_V=I$ and
\begin{align}
\label{eq:lcu_operators_chebyshev}
    A_V &= \sum_{j=0}^{n-1} \sqrt{\frac{2^j}{2^n-1}} \dyad{j}{0}^{\otimes m} \otimes I^{\otimes n} + \text{ u.c. },\\
    B_V &= \sum_{j=0}^{n-1} \dyad{j} \otimes X_jZ_jX_j.
\end{align}
Such a block-encoding requires $m=b = \lceil \log_2 n \rceil$ ancilla qubits.

\begin{figure*}[t]
\centering
\begin{subfigure}[t]{.65\linewidth}
(a) \quad
\begin{quantikz}[row sep={0.7cm, between origins}, column sep=0.2cm]
    & \ctrl{1}  & \qw \\
    & \gate[2]{U}  &\qw\\
    & \qw & \qw
\end{quantikz}
 =
\begin{quantikz}[row sep={0.7cm, between origins}, column sep=0.19cm]
    & \ctrl[style={NormalRed},wire style={dotted}]{1} & \ctrl{1} & \ctrl{1} & \ctrl[style={NormalRed},wire style={dotted}]{1}  & \ctrl{1} & \qw \\
    & \gate{A}  & \gate[2]{B} & \gate{C} & \gate{A^\dagger} &  \gate{R} & \qw\\
    & \qw & \qw & \qw & \qw & \qw & \qw
\end{quantikz}
\end{subfigure}
\vspace{.5cm}
\vfill
\begin{subfigure}[t]{.65\linewidth}
(b) \quad
\begin{quantikz}[row sep={0.7cm, between origins}, column sep=0.2cm]
     & \ctrl{1}  & \qw \\
     & \gate[2]{U^2}  &\qw\\
     & \qw & \qw
\end{quantikz}
 =
\begin{quantikz}[row sep={0.75cm, between origins}, column sep=0.19cm]
    & \qw & \ctrl[style={NormalBlue},wire style={dotted}]{1} & \ctrl[style={NormalBlue},wire style={dotted}]{1} & \qw  & \ctrl[style={NormalYellow},wire style={dotted}]{1} & \qw & \ctrl[style={NormalBlue},wire style={dotted}]{1} & \ctrl[style={NormalBlue},wire style={dotted}]{1} & \qw  & \ctrl[style={NormalYellow},wire style={dotted}]{1} & \qw\\
    & \gate{A}  & \gate[2]{B} & \gate{C} & \gate{A^\dagger} &  \gate{R} & \gate{A}  & \gate[2]{B} & \gate{C} & \gate{A^\dagger} &  \gate{R} & \qw\\
    & \qw & \qw & \qw & \qw & \qw & \qw & \qw & \qw & \qw & \qw & \qw
\end{quantikz}
\end{subfigure}
\caption{Simplifications used in the block-encoding circuit of Chebyshev basis functions. (a) For controlled-$U$, where $U$ is any LCU circuit, one can remove the controls depicted in red. (b) For the specific case where $B^2=I$ and $C^2=I$ it is possible to simplify controlled-$U^2$ as well. One can remove either the blue or the yellow colored controls, depending on which option is more convenient. Our method uses controlled-$U^{2^i}$ and thus benefits significantly from these techniques.}
\label{fig:controlled_LCU}
\end{figure*}
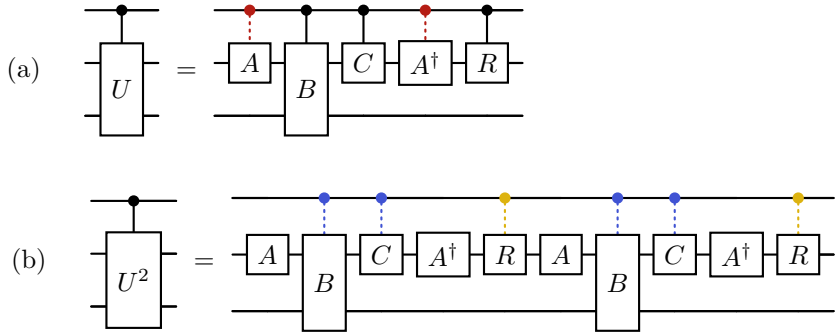

Next, Lemma 8 in Ref.~\cite{Low_2019} shows that a block-encoding $V$ can be used to derive another block-encoding $U_V$ that satisfies the qubitization condition
\begin{align}
\label{eq:qubitization1}
    U_V = \bigoplus_\xi \begin{pmatrix} \xi & -\sqrt{1- \xi^2} \\ \sqrt{1- \xi^2} &  \xi \end{pmatrix}_\xi ,
\end{align}
where $\xi$'s are the eigenvalues of the block-encoded marix and each term is represented in the basis $\left\{ |G_\xi\rangle, |G_\xi^\bot\rangle \right\}$ defined in Ref.~\cite{Low_2019}. Moreover, Corollary 9 in Ref.~\cite{Low_2019} shows that in the special case where $V^2 = I$ (which holds true here given that $B_V^2=I$ and $C_V=I$) the qubitized block-encoding takes a simple form
\begin{align}
\label{eq:qubitization2}
    U_V = \left( \big( 2\dyad{0}^{\otimes b} - I^{\otimes b} \big) \otimes I^{\otimes n} \right) V .
\end{align}

We now combine two results from the literature. The first result is Lemma 16 in Ref.~\cite{Childs_2017}. It states that for positive integer $k$
\begin{multline}
    \begin{pmatrix} \xi & -\sqrt{1- \xi^2} \\ \sqrt{1- \xi^2} &  \xi \end{pmatrix}^k = \\ \begin{pmatrix} T_k(\xi) & -\sqrt{1- \xi^2} \; u_{k-1}(\xi) \\ \sqrt{1- \xi^2} \; u_{k-1}(\xi) &  T_k(\xi) \end{pmatrix},
\end{multline}
where $u_k$ are Chebyshev polynomials of the second kind. For $k=0$ we define the above expression as the identity matrix. The second result we use is Lemma 54 in Ref.~\cite{Gilyen_2019} which deals with products of block-encodings. In the case where the encoded matrices are unitaries and their normalization constant is $\lambda=1$, the product of $(1,b,0)$-block-encodings yields the $(1,b,0)$-block-encoding of the product. Moreover, since $U_V$ is the direct sum in Eq.~\eqref{eq:qubitization1}, powers of $U_V$ preserve the qubitization condition. It follows that
\begin{align}
\label{eq:cheb_operator}
    U^k_V& = \bigoplus_\xi \begin{pmatrix} T_k(\xi) & -\sqrt{1- \xi^2} \; u_{k-1}(\xi) \\ \sqrt{1- \xi^2} \; u_{k-1}(\xi) &  T_k(\xi) \end{pmatrix}_\xi,
\end{align}
and $U^0_V = I$. This is a $(1, b, 0)$-block-encoding of the $k$-th Chebyshev polynomial applied to $H^C$, i.e. $T_k(H^C) = (\bra{0}^{\otimes b} \otimes I^{\otimes n}) U_V^k (\ket{0}^{\otimes b} \otimes I^{\otimes n})$. Note that the algebra of qubitized block-encodings is also used in Ref.~\cite{Kirby2023exactefficient} to implement Krylov basis vectors and Lanczos' method on a quantum computer. 
We emphasize that our method works because $\lambda= \| H^C \| = 1$. For $\lambda \neq 1$ we would not obtain the correct result after rescaling the eigenvalues of $H$ since $T_k(H/\lambda) \neq T_k(H)/\lambda $ (but see Section 5.1 of Ref.~\cite{Camps2022} for approximate solutions when this issue arises). 

To construct the finite Chebyshev series Eq.~\eqref{eq:f_d_Chebyshev} from the block-encoding circuit $U^k_V$ we insert $U^k_V$ into Eq.~\eqref{eq:lcu_operator_B}. To  implement $U^k_V$ controlled on the value of $k$ we again use the identity in Eq.~\eqref{eq:uid} with $m=a$ and obtain 
\begin{equation}
\label{eq:ctrl_qubitization}
\begin{split}
   B^C&=\sum_{k=0}^{K-1} \dyad{k}\otimes U^k_V\\
   &= \prod_{i=0}^{a-1} \left[ \dyad{0}_i \otimes I^{\otimes n+b} + \dyad{1}_i \otimes U^{2^i}_V \right].
\end{split}
\end{equation}
We can save some resources when controlling $U_V$. First, we can remove the control on $A_V$ and $A_V^\dag$ since they cancel when no operation is performed in between. Second, when controlling $U_V^2$, we can either remove the control on $B_V$ or on $R=2\dyad{0}^{\otimes b} - I^{\otimes b}$. These improvements are shown in the circuit diagrams in Fig.~\ref{fig:controlled_LCU} and for our purposes we remove the control on $B_V$.

To provide a gate count for $B^C$ in Eq.~\eqref{eq:ctrl_qubitization}, we start with a gate count for $\dyad{1}_i \otimes U_V^2$ recalling $U_V=RA_V^\dagger C_VB_VA_V$. The operators $A_V$ and $A_V^\dagger$ in $U_V^2$ can be implemented with a $b$-qubit state preparation circuit with real coefficients, requiring $2^b$ \CX gates each \cite{Bergholm_2005}. $B_V$ costs $n$ $b$-qubit controlled $Z$ gates, which we denote as \CbZ. Controlled $R$ costs one \CbZ gate. In total $\dyad{1}_i \otimes U_V^2$ costs $2(n + 1)$ \CbZ gates and $2\cdot2^{b}$ \CX gates. For Eq.~\eqref{eq:ctrl_qubitization} we need to implement $2^{i-1}$ copies of $\dyad{1}_i \otimes U_V^2$. By summing up the gate counts for each copy we get the following totals for $B^C$: $\sum_{i=0}^{a-1} 2^i (n+1) = (2^a-1)(n+1)$ \CbZ gates and $\sum_{i=0}^{a-1} 2^i\cdot 2^{b}= (2^a-1)2^b$ \CX gates.

We wish to reduce gate counts to two-qubit gates only. Each \CbZ gate can be implemented with $\mathcal{O}(b)$ gates \cite{Nie_2024}, giving a total for $B^C$ of $\mathcal{O}\left[ \left( 2^a - 1\right) \left( (n + 1) b + 2^b \right)\right] = \mathcal{O}(2^a n b)$ \CX gates. Here we kept only dominant terms in $d$ and $n$. 

The best choice decompositions of \CbZ will depend on $b$ and the resources available, with alternative \CbZ decompositions, such as in ~\cite{daSilva_2022}, or strategies for synthesising series of \CbZ with varying control conditions, such as with the unary iteration circuits proposed in ~\cite{Babbush_2018}, potentially providing reduced \CX count.

\begin{figure*}[t]
\centering
    \begin{quantikz}[row sep={0.7cm,between origins}]
        \lstick[3]{ancillas for\\coefficients $\ket{0}^{\otimes a}$} 
        & \gate[3]{A} 
        & \ctrl{3} \gategroup[wires=9,steps=4,style={dotted,cap=round,inner sep=2pt}]{$B$} 
        & 
        & \quad \dots\quad
        & 
        & \gate[3]{C} 
        & \gate[3]{A^\dag} 
        & \rstick[3]{$\bra{0}^{\otimes a}$}
        \\
        & 
        & 
        & \ctrl{2} 
        & \quad \dots\quad 
        & 
        & 
        & 
        &
        \\
        & 
        & 
        & 
        & \quad \dots\quad 
        & \ctrl{1} 
        & 
        & 
        & 
        \\
        \lstick[2]{ancillas for\\block-encoding $\ket{0}^{\otimes b}$
        \\ (Chebyshev only)} 
        & 
        & \gate[6]{U^{2^0}} 
        & \gate[6]{U^{2^1}} 
        & \quad \dots\quad 
        & \gate[6]{U^{2^{a-1}}} 
        & 
        & 
        & \rstick[2]{$\bra{0}^{\otimes b}$}
        \\
        & & & & \quad \dots\quad & & & &
        \\
        \lstick[4]{main register $\ket{+}^{\otimes n}$} & & & & \quad \dots\quad & & & 
        & \rstick[4]{$\propto \sum_x f_d(x)\ket{x}$}
        \\
        & & & & \quad \dots\quad & & & &
        \\
        & & & & \quad \dots\quad & & & &
        \\
        & & & & \quad \dots\quad & & & &
    \end{quantikz}
    \caption{Quantum circuit for state preparation in one dimension. The amplitudes are defined by a finite series, such as Fourier or Chebyshev. Circuit $U$ is the block-encoding of the corresponding basis functions and may require $b$ additional ancilla qubits. Each bit in the coefficient register drives the application of powers of the block-encoding circuit. For example, in the subspace where the coefficient ancillas are in state $\ket{k}$, we apply $U^k$ to the other two registers. The algorithm succeeds if we measure the zero state on all $a + b$ ancilla qubits.
    }
    \label{fig:quantum_circuit_1d}
\end{figure*}
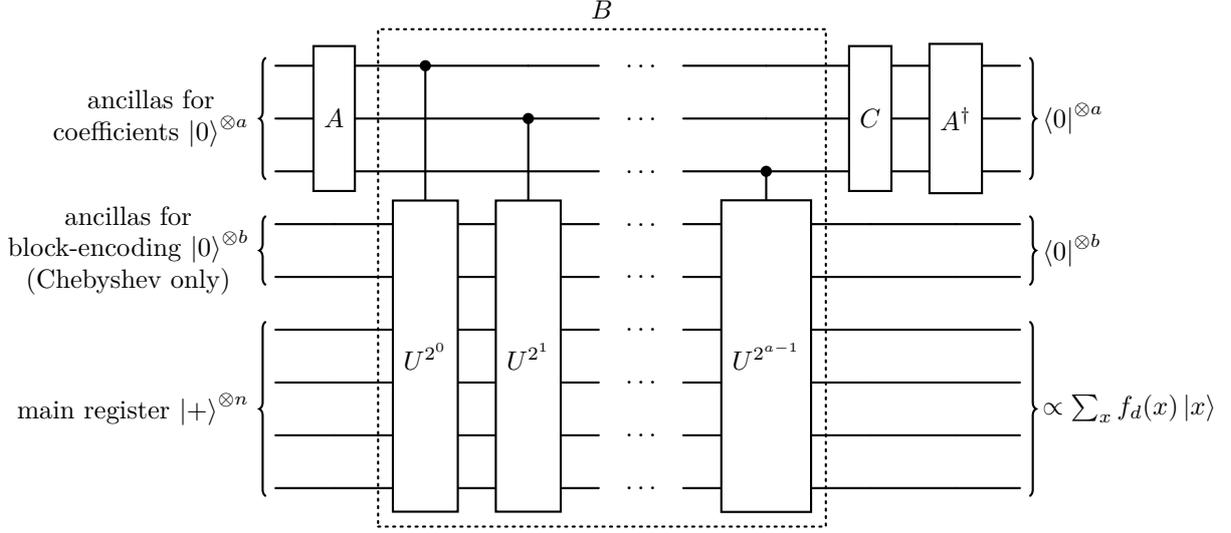

\subsection{State preparation in 1 and 2 dimensions}

We now utilize our building-blocks for the construction of quantum states defined by truncated series expansions. In this section we drop the superscripts $F$ for Fourier and $C$ for Chebyshev, since the protocols are almost identical. Then, the $n$-qubit discretization of the domain ($2^n$ points) is denoted as $H$, and the block-encoding of the $k$-th basis function is denoted as $U_k$.

We begin with one-dimensional functions $f$ approximated by either Eq.~\eqref{eq:f_d_Fourier} or Eq.~\eqref{eq:f_d_Chebyshev}, here both denoted by $f_d$. Without further assumptions, the number of terms is $K=2d+1$ for Fourier series and $K=d+1$ for Chebyshev series. Each basis function is associated to a complex coefficient $c_k$ which we rewrite as $c_k = \e^{\im \phi_k} |c_k|$. Then
\begin{align}
f_d(H) = \sum_{k=0}^{K-1} c_k U^k 
     = \mathcal{N} \sum_{k} \e^{\im \phi_k} \frac{|c_k|}{\mathcal{N}} U^k ,
\end{align}
where we normalize the coefficients with $\mathcal{N}= \sum_{k} |c_k|$. We encode the coefficients using operators $A$ and $C$ (Eqs.~\eqref{eq:lcu_operator_A} and~\eqref{eq:lcu_operator_C}), and an ancilla register of size $a=\lceil\log_2 K\rceil$. We then use the operator $B$ defined in Eq.~\eqref{eq:fourier_b} for Fourier and Eq.~\eqref{eq:ctrl_qubitization} for Chebyshev noting, however, that the Chebyshev method requires additional $b$ ancilla qubits. We thus obtain a $(\mathcal{N},a+b,0)$-block-encoding $A^\dag C B A$ with
\begin{multline}
    \left(\bra{0}^{\otimes (a + b)} \otimes I^{\otimes n} \right) \; A^\dag C B A \; \left( \ket{0}^{\otimes (a + b)} \otimes I^{\otimes n} \right) =\\
    \frac{1}{\mathcal{N}} f_d(H) . 
\end{multline}
We apply $A^\dag C B A$ to $\ket{0}^{\otimes(a+b)} \otimes \ket{+}^{\otimes n}$ with the uniform superposition $\ket{+}^{\otimes n} = \frac{1}{\sqrt{2^n}}\sum_{x} \ket{x}$ on the main register. 
Upon measuring the zero state for all the ancilla qubits, the protocol generates the unnormalized state $\frac{1}{\mathcal{N} \sqrt{2^n}} \sum_{x} f_d(x) \ket{x}$. The probability of success is given by $p_{\mathrm{success}} =  \frac{\sum_x |f_d(x)|^2}{\mathcal{N}^2 2^n}$. Renormalizing the state (dividing by $\sqrt{p_\mathrm{success}}$) we obtain the desired result $\sum_x g_d(x) \ket{x} := \frac{1}{\sqrt{\sum_x |f_d(x)|^2}} \sum_{x} f_d(x) \ket{x}$. The full circuit is sketched in Fig.~\ref{fig:quantum_circuit_1d}.

\begin{figure*}[t]
\centering
    \begin{quantikz}[row sep={0.75cm, between origins}]
    \lstick{$\ket{0}$} & 
        \qwbundle{a_x} & 
        \gate[2]{A} & 
        \ctrl{2} \gategroup[wires=4,steps=2,style={dotted,cap=round,inner sep=0pt},label style={yshift=-1pt}]{$B_X$} & 
        \quad \dots\quad & 
        & 
        & 
        \gate[2]{C} & 
        \gate[2]{A^\dag} & \\
    \lstick{$\ket{0}$} & 
        \qwbundle{a_y} & 
        &  
        & 
        & 
        \ctrl{3} \gategroup[wires=5,steps=2,style={dotted,cap=round,inner sep=0pt},label style={yshift=-2pt}]{$B_Y$} & 
        \quad \dots\quad &
        & 
        & \\
    \lstick{$\ket{0}$} & 
        \qwbundle{b_x} & 
        & 
        \gate[2]{U} & 
        \quad \dots\quad &
        & 
        &
        &  
        & \\
    \lstick{$\ket{+}$} & 
        \qwbundle{n_x} & 
        & 
        & 
        \quad \dots\quad & 
        & 
        &
        &  
        & \\
    \lstick{$\ket{0}$} & 
        \qwbundle{b_y} & 
        & 
        & 
        & 
        \gate[2]{U} & 
        \quad \dots\quad &
        & 
        & \\
    \lstick{$\ket{+}$} & 
        \qwbundle{n_y} & 
        & 
        & 
        & 
        &
        \quad \dots\quad  &
        & 
        &
    \end{quantikz}
    \caption{Quantum circuit for state preparation in two dimensions.}
    \label{fig:quantum_circuit_2d}
\end{figure*}
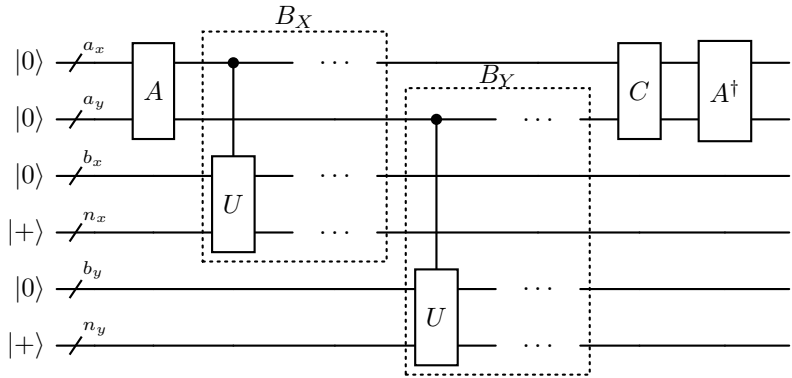

With the one-dimensional warm up, we have all the elements to construct states defined by bivariate functions $f(x, y)$. Again we start by discretizing the variables using $n_x$ and $n_y$ qubits in the main registers, which gives us a two-dimensional grid of $2^{n_x} \times 2^{n_y}$ points. We then approximate $f$ as a finite series of degrees $d_x$ and $d_y$.
The coefficients are rewritten as $c_{k,l} = \e^{\im \phi_{k,l}}|c_{k,l}|$ where $k=0, \dots, K_x-1$ and $l=0, \dots, K_y-1$, and $K_x$, $K_y$ depend on the series type and degree as before. The normalization constant becomes $\mathcal{N}=\sum_{k} \sum_{l} |c_{k,l}|$. The LCU operators are now defined as:
\begin{widetext}
\begin{align}
    A &= \sum_{k=0}^{K_x-1} \sum_{l=0}^{K_y-1} \sqrt{\frac{|c_{k,l}|}{\mathcal{N}}} \dyad{k}{0}^{\otimes {a_x}} \otimes \dyad{l}{0}^{\otimes {a_y}} \otimes I^{\otimes (b_x + b_y + n_x + n_y)} + \text{ u.c. },\\
    B_X &= \sum_{k=0}^{K_x-1} \dyad{k} \otimes I^{\otimes a_y} \otimes U^k \otimes I^{\otimes (b_y + n_y)}, \\
    B_Y &= \sum_{l=0}^{K_y-1} I^{\otimes a_x} \otimes \dyad{l} \otimes I^{\otimes (b_x + n_x)} \otimes U^l, \\
    C &= \sum_{k=0}^{K_x-1} \sum_{l=0}^{K_y-1} \e^{\im \phi_{k,l} } \dyad{k} \otimes \dyad{l} \otimes I^{\otimes (b_x + b_y + n_x + n_y)}.
\end{align}
Applying the circuit to the uniform superposition of both main registers and post-selecting on the zero state of all the ancillas yields the desired state
\begin{equation}
\begin{split}
\label{eq:unnorm_2d_state}
    \left(\bra{0}^{\otimes (a_x + a_y + b_x +b_y)} \otimes I^{\otimes (n_x + n_y)} \right) A^\dag C B_Y B_X A \left( \ket{0}^{\otimes (a_x + a_y + b_x +b_y)} \otimes 
    \ket{+}^{\otimes (n_x + n_y)} \right) =\\
    \frac{1}{\mathcal{N}\sqrt{2^{n_x+n_y}}} \sum_x \sum_y f_d(x, y) \ket{x} \otimes \ket{y}.
\end{split}
\end{equation}
\end{widetext}

The protocol succeeds with probability 
\begin{align}\label{eq:success-analytical}
    p_{\mathrm{success}} =  \frac{\sum_x \sum_y |f_d(x,y)|^2}{\mathcal{N}^2 2^{n_x + n_y}} .
\end{align}
To obtain the desired result we renormalize Eq.~\eqref{eq:unnorm_2d_state} by dividing by $\sqrt{p_\text{success}}$. The full circuit for bivariate state preparation is sketched in Fig.~\ref{fig:quantum_circuit_2d}. The generalization of this protocol to arbitrary number of variables is straightforward.

\subsection{Resource scaling for state preparation in arbitrary dimension}\label{sec:scaling}

We can extrapolate the resource costs given in the previous sections for the one-dimensional case to estimate the scaling of state preparation in arbitrary dimensions. We focus on the number of qubits and \CX gate count as a proxy for the two-qubit gate count, which is the main resource cost for our algorithm on near-term hardware. We neglect single-qubit gates. 

Let us recall the resource requirements for state preparation with one variable ($D=1$). The operator $B$ costs $\mathcal{O} (an)$ \CX gates for the Fourier approach (Sec.~\ref{sec:fourier}) and $\mathcal{O}\left( 2^a n b \right)$ \CX gates for the Chebyshev approach (Sec.~\ref{sec:chebyshev}). The size of the ancilla register $a$ is determined by the number of coefficients in the approximating series $f_d$, namely, $a=\lceil \log_2(2d+1) \rceil$ for Fourier series and $a=\lceil \log_2(d+1)\rceil$ for Chebyshev series. The block-encodings of the Chebyshev basis functions require an additional ancilla register of size $b=\lceil \log_2 n \rceil$. $A$, $A^\dagger$, Eq.~\eqref{eq:lcu_operator_A}, and $C$, Eq.~\eqref{eq:lcu_operator_C}, can be implemented with $a$-qubit circuits requiring $\mathcal{O}\left(2^a\right)$ \CX gates each.

\begin{table*}[tb]
    \centering
    \def\arraystretch{1.5}
    \begin{tabular}{c|cc}
    \toprule
        & Fourier
        & Chebyshev\\
     \midrule
     Target function
     & $[0,1]^D \rightarrow \mathbb{C}$ 
     & $[-1, 1]^D \rightarrow \mathbb{C}$\\
     Classical preprocessing
     &  \multicolumn{2}{c}{Cost of computing coefficients, e.g. $\mathcal{O}(D d^D \log d)$} for FFT\\
     Main qubits
     & $Dn$
     & $Dn$ \\
     Ancilla qubits
     & $D \lceil \log_2 (2d +1)\rceil$
     & $D \lceil \log_2 (d+1)\rceil + D\lceil\log_2 n\rceil$\\
     Two-qubit gates
     & $\mathcal{O}(d^D + Dn \log d)$
     & $\mathcal{O}(d^D + Ddn\log n)$\\
     Success probability
     & $\mathcal{N}^{-2} 2^{-Dn} \sum_{\bm{x}} |f_d(\bm{x})|^2$
     & $\mathcal{N}^{-2} 2^{-Dn} \sum_{\bm{x}} |f_d(\bm{x})|^2$\\
     \bottomrule
    \end{tabular}
    \caption{Resources of our algorithm for preparing a quantum state with amplitudes proportional to a $D$-dimensional function represented on an equidistant grid of total size $2^{Dn}$. The method uses a Fourier or Chebyshev polynomial approximation $f_d$ with maximal degree $d$ in each dimension. $\mathcal{N}=\sum_{\bm{k}} |c_{\bm{k}}|$ is a normalization dependent on the expansion coefficients $c_{\bm{k}}$.}
    \label{tab:resources}
\end{table*}

We now provide the resource requirements for state preparation in arbitrary dimension. First, we assume that there are $D$ variables, each discretized with $n$ qubits, i.e. a total of $Dn$ qubits in the main registers. Second, we assume the same degree $d$ for each variable. The coefficients are encoded in ancilla registers $a_1, a_2, \dots, a_D$ each with size $a$. Then the total number of ancilla qubits is $Da=D\lceil \log_2(2d+1) \rceil$ and $D(a+b)=D(\lceil \log_2(d+1)\rceil + \lceil \log_2 n\rceil)$ for the Fourier and Chebyshev approaches, respectively, where we inserted the respective expressions from the 1D case for $a$ and $b$. The qubit counts are summarized in Table~\ref{tab:resources}.

In general, the coefficients in a multivariate approximation $f_d$ may take arbitrary values. Then the  operators $A$, $A^\dagger$ and $C$ can be implemented with $Da$-qubit circuits requiring $\mathcal{O}\left(2^{Da}\right)$ \CX gates each. We have $D$ operators $B_1, B_2, \dots, B_D$. Each $B_i$ is controlled only on the corresponding coefficient ancilla register $a_i$ and acts only on the corresponding main register $n_i$. For the Chebyshev case each $B_i$ also acts on the corresponding block-encoding ancilla register $b_i$, cf. Fig.~\ref{fig:quantum_circuit_2d}. Hence, implementing all $B_i$ is $D$ times the 1D implementation cost of $B$: $\mathcal{O} (Dan)$ \CX gates for the Fourier approach and $\mathcal{O} \left(D 2^a n b\right)$ \CX gates for the Chebyshev approach. Summing the costs for $A$, $A^\dagger$, $C$ and all $B_i$ and inserting the values for $a$ and $b$ yields $\mathcal{O}\left(2^{D\lceil \log_2(2d+1)\rceil} + Dn\lceil \log_2(2d+1)\rceil\right)$ \CX gates for the Fourier approach and $\mathcal{O}\left(2^{D\lceil \log_2(d+1)\rceil} + D n\lceil \log_2 n \rceil 2^{\lceil \log_2(d+1)\rceil} \right)$ \CX gates for the Chebyshev approach. The simplified expressions $\mathcal{O}\left(d^D + Dn\log d\right)$ and $\mathcal{O}\left( d^D + Ddn\log n\right)$, respectively, are 
provided as the two-qubit gate count in Table~\ref{tab:resources}. Appendix~\ref{app:prior_works} and Table~\ref{tab:prev_work} compare our resource requirements to those of other state preparation methods for multivariate functions.

Some remarks are in order. First, if we wanted to boost the probability of success to unity, we could use amplitude estimation. That would increase the number of gates by a multiplicative factor of $\mathcal{O}\left(p_\text{success}^{-1/2}\right)$. Second, the degree $d$ for each dimension must be chosen depending on the desired accuracy $\varepsilon$. The scaling of $d(\varepsilon)$ is problem-dependent and is not analyzed in this work. However, the protocol is efficient only if $d \in \mathcal{O}(\mathrm{poly}(n))$, which limits the accuracy that can be achieved in general. Third, any method that implements a discretized version of the target function must rely on a grid of points. In our case, each qubit added to a main register increases the grid resolution exponentially, while increasing the number of two-qubit gates only mildly.

\section{Results}

\begin{figure}[t]
    \centering
    \includegraphics[width=\linewidth]{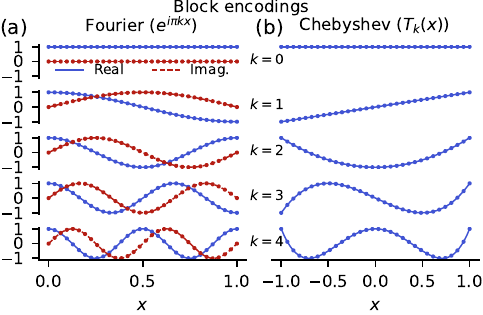}
    \caption{Block-encodings of the Fourier and Chebyshev basis functions. Markers are the diagonal of the Fourier block encoding, Eq.~\eqref{eq:rte_operator} in (a), and $(\bra{0}^{\otimes b} \otimes I^{\otimes n}) U^k (\ket{0}^{\otimes b} \otimes I^{\otimes n})$ with Chebyshev block encoding $U^k$, Eq.~\eqref{eq:cheb_operator} in (b), for $n=5$ qubits (32 grid points). Lines are the corresponding continuous basis functions $\e^{\im\pi k x}$ and $T_k(x)$.}
    \label{fig:block_encodings}
\end{figure}

With our first numerical simulations we visually inspect the block-encodings of Fourier and Chebyshev basis functions. In Fig.~\ref{fig:block_encodings}(a) we plot the diagonal of Eq.~\eqref{eq:rte_operator} for $k=0,\dots, 4$. In Fig.~\ref{fig:block_encodings}(b) we show the diagonal of $(\bra{0}^{\otimes b} \otimes I^{\otimes n}) U^k (\ket{0}^{\otimes b} \otimes I^{\otimes n})$ where $U^k$ is given in Eq.~\eqref{eq:cheb_operator}. In both cases we use $n = 5$, leading to a $x$-axis discretization with $32$ points. As expected the data points closely track $\e^{\im k \pi H}$ in Fig.~\ref{fig:block_encodings}(a), and $T_k(H)$ in Fig.~\ref{fig:block_encodings}(b).

\subsection{2D Ricker wavelet via Chebyshev series}

To illustrate the Chebyshev approach, we consider a 2D Ricker wavelet centered at $(0, 0)$,
\begin{equation}
    f(x, y) = \frac{1}{\pi\sigma^4} \left( 1 - \frac{x^2 + y^2}{2\sigma^2} \right) \e^{-\frac{x^2 + y^2}{2\sigma^2}}.
\end{equation}
Wavelets of this form are used, e.g. in image or seismic data applications~\cite{wang2015generalized,antoineWaveletAnalysisSignals2000,wrightNoisyIntermediatescaleQuantum2024}.

\begin{figure*}[t]
    \centering
    \includegraphics[width=\textwidth]{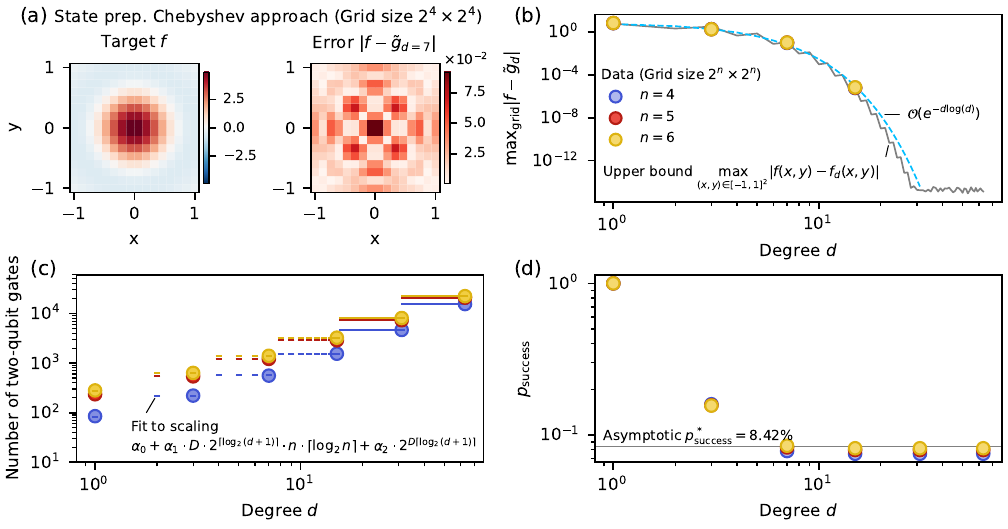}
    \caption{Resources for preparing an approximation to the 2D Ricker wavelet with products of Chebyshev polynomials of degree $d$ in each dimension. (a) Discretized target function and the absolute difference to the rescaled amplitudes from a noiseless simulation of the algorithm for $d=7$. (b) Maximum absolute error of the prepared state evaluated at the grid points for different grid sizes. The difference between the prepared state and the Chebyshev approximation is at machine precision (not shown). Hence, the maximum error of the noiseless simulation against the target function is upper bounded by the maximum error of the degree-$d$ Chebyshev interpolant over the (non-discretized) domain (grey line -- floored at machine precision). The dashed line illustrates a super-exponential asymptotic error bound $\mathcal{O}\left(\e^{-d\log d}\right)$. (c) Two-qubit gate count for different grid sizes compiled with \texttt{pytket} optimization level 2 for the H2-1 default gate set. The line markers are a least-square fit of the data (round markers) to the worst-case scaling with $\alpha_0=-29.3, -47.3, -49.3$, $\alpha_1=3.14, 4.47, 4.39$, $\alpha_2=3.00, 3.00, 3.00$ for $n=4, 5, 6$ and $D=2$. (d) Empirical (markers) and asymptotic success probability for $n\rightarrow\infty$ and $d\rightarrow\infty$ (line).}
    \label{fig:Ricker-chebyshev-results}
\end{figure*}

We interpolate the function on the domain $[-1, 1]^2$ with the Chebyshev series Eq.~\eqref{eq:f_d_Chebyshev_2d} and compute its coefficients via the fast Fourier transform (see Appendix~\ref{app:Chebyshev}). We choose the same degree and same grid size in each variable, i.e. $d_x=d_y=d$ and $n_x=n_y=n$. 
Figure~\ref{fig:Ricker-chebyshev-results} shows results for approximating the 2D Ricker wavelet with $\sigma=0.5$ by a Chebyshev series of varying degree $d$ and for different grid sizes $2^n \times 2^n$. Figure~\ref{fig:Ricker-chebyshev-results}(a) illustrates the discretized target function and its absolute difference to the amplitudes of the prepared quantum state for degree $d=7$ and grid size $2^4\times 2^4$. For this comparison we rescale the amplitudes from a noiseless simulation of the algorithm appropriately,
\begin{equation}
\label{eq:tilde_g_d}
    \tilde{g}_d(x,y) = g_d(x,y) \sqrt{\sum_{i=0}^{2^n-1} \sum_{j=0}^{2^n-1} |f(x_i, y_i)|^2},
\end{equation}
where $x_i, y_i$ are the evaluated grid points. Figure~\ref{fig:Ricker-chebyshev-results}(b) shows the maximum error $\max_{(x, y)} |f(x, y) - \tilde g_d(x, y)|$ where the maximum is over grid points. Increasing the degree leads to rapid convergence of the amplitudes to the target function. We observe that the error between the rescaled amplitudes and the Chebyshev approximation $f_d$ is at machine precision (not shown). Hence, the convergence of our method to the target function is determined by the convergence of the Chebyshev approximation. The grey line shows the maximum error of the Chebyshev approximation, $\max_{(x,y) \in [-1,1]^2} |f(x,y) - f_d(x, y)|$, evaluated numerically over the whole domain (not just the grid) via global optimization. The line is consistent with supergeometric convergence to the target function. The numerics are floored at machine precision from around degree 30. The data from the algorithm's simulation is indeed upper bounded by this maximum error demonstrating that the accuracy tracks the Chebyshev approximation error. The dashed line illustrates a supergeometric bound $\mathcal{O}\left(\e^{-d\log d}\right)$. This indicates that the accuracy $\varepsilon$ is bounded by $\mathcal{O}\left(\e^{-d\log d }\right)$. Hence, we can choose degree $d = \mathcal{O}\left(\e^{W_0[\log(1/\varepsilon)]}\right) = \mathcal{O}\left[\log(1/\varepsilon)\right]$ for a fixed target error $0 < \varepsilon < 1$ with $W_0$ the principal branch of the Lambert W function. As a rule-of-thumb we expect at least geometric convergence to hold for analytic functions.

Figure~\ref{fig:Ricker-chebyshev-results}(c) shows the number of two-qubit gates for varying degrees and different grid sizes. Circuits are compiled with \texttt{pytket} for the default native gate set of the H2-1 quantum computer at optimization level $2$~\cite{sivarajahVertKetRangle2020}. The default native two-qubit gate is $\RZZ(\theta)=\e^{-\im\frac{\theta}{2}Z\otimes Z}$ with arbitrary angle $\theta$. We perform a least-square fit of the data to the scaling of two-qubit gates derived in Sec.~\ref{sec:scaling}, $\alpha_0 + \alpha_1 D 2^{\lceil \log_2(d+1) \rceil} n\lceil \log_2 n \rceil + \alpha_2 2^{D\lceil \log_2(d+1) \rceil}$, where $\alpha_0, \alpha_1, \alpha_2$ are the fitted parameters and we added a constant offset. The fit in Fig.~\ref{fig:Ricker-chebyshev-results}(c) shows good agreement. For large $n$ and $d$ this scaling law is upper bounded by the simplified expression in Table~\ref{tab:resources}. The best parameters from the fit (see caption of Fig.~\ref{fig:Ricker-chebyshev-results}) are in line with expectations from a worst-case analysis. For example, the term $\alpha_2 2^{D\lceil \log_2(d+1) \rceil}$ comes from the implementation of the operators $A$, $C$, $A^\dagger$ in Fig.~\ref{fig:quantum_circuit_2d} and the value $\alpha_2=3$ indicates that each of the operators is implemented with the expected worst-case cost $2^{D\lceil \log_2(d+1) \rceil}$. We note some residual variance in the coefficients $\alpha_0$ and $\alpha_1$ depending on $n$, which we attribute to suppressed, non-dominant terms and compiler optimizations. The total number of qubits can be computed from Table~\ref{tab:resources}. For example, the largest circuit in Fig.~\ref{fig:Ricker-chebyshev-results}(c) at $d=63$, $n=6$ requires 30 qubits (12 for the main register and 18 for ancilla registers). The success probability shown in Fig.~\ref{fig:Ricker-chebyshev-results}(d) saturates at around $8\%$, which is close to the asymptotic success probability for $n\rightarrow\infty$, $d\rightarrow\infty$. We compute the asymptotic success probability as $p_\text{success}^* = \int_{[-1,1]^2}\diff x\diff y |f(x, y)|^2 / 4 \big(\sum_{k,l=0}^{30} |c_{k,l}|\big)^2$, which is the continuous version of Eq.~\eqref{eq:success-analytical}. The sum in the numerator runs to $k,l=30$ because higher-order coefficients do not contribute noticeably and fall off super-exponentially. The factor $1/4$ is from the size of the integration domain.

\subsection{Bivariate Student's t-distribution via Fourier series}
\label{ssec:student-t}

To illustrate the Fourier approach, we consider the bivariate non-standardized Student's t-distribution with one degree of freedom ($\nu=1$, this is also known as the bivariate Cauchy distribution)
\begin{align}
    f(\bm{x}) =  \frac{1}{2 \pi \sqrt{\det(\Sigma)}} \left[1 + (\bm{x} - \bm{\mu})^T \Sigma^{-1}(\bm{x} - \bm{\mu}) \right]^{-\frac{3}{2} } ,
\end{align}
where $\bm{x}=(x, y)^T$ and $\bm{\mu}$ and $\Sigma$ are location and scale parameters, respectively. Student's t-distribution is characterized by heavy tails with applications in finance to model heavy-tailed return distributions~\cite{chiuForecastingVARModels2017} among others.

\begin{figure*}[t]
    \centering
    \includegraphics[width=\textwidth]{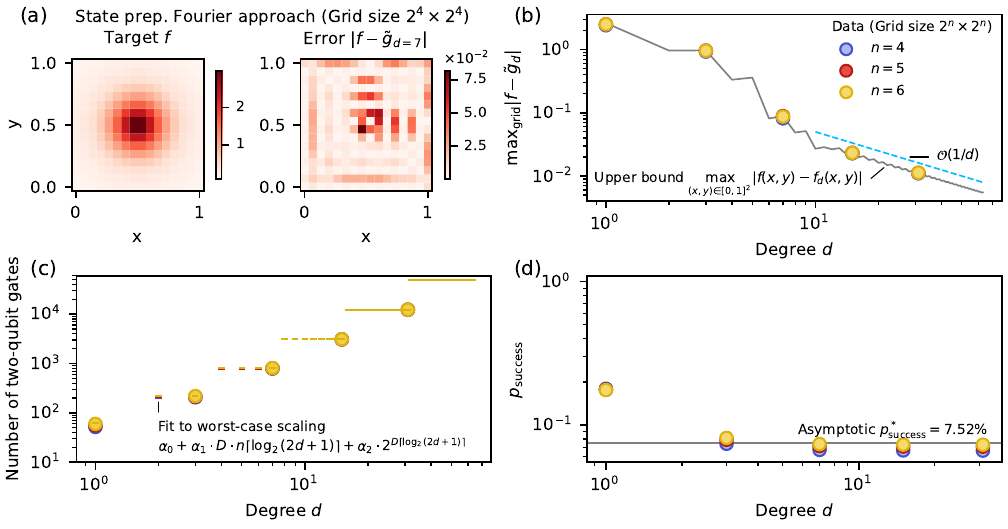}
    \caption{Resources for preparing an approximation to the bivariate Student's t-distribution with a 2D Fourier series approximation of degree $d$ in each dimension. (a) Discretized target function and the difference to the rescaled amplitudes from a noiseless simulation of the algorithm for $d=7$. (b) Maximum absolute error of the prepared state evaluated at the grid points for different grid sizes. The difference between the prepared state and the Fourier approximation is at machine precision (not shown). Hence, the maximum error of the noiseless simulation against the target function is upper bounded by the maximum error of the maximal degree-$d$ Fourier interpolant over the (non-discretized) domain (grey line -- floored at machine precision). The dashed line illustrates the asymptotic error bound $\mathcal{O}(1/d)$. (c) Two-qubit gate count of the algorithm for different grid sizes compiled with \texttt{pytket} optimization level 2 for the H2-1 gate set. The line markers are a least-square fit of the data (round markers) to the worst-case scaling with $\alpha_0=-4.91, -4.91, -4.91$, $\alpha_1=0.78, 0.82, 0.85$, $\alpha_2=2.99, 2.99, 2.99$ for $n=4, 5, 6$ and $D=2$. (d) Empirical (markers) and asymptotic success probability for $n\rightarrow\infty$ and $d\rightarrow\infty$ (line).}
    \label{fig:Cauchy-fourier-results}
\end{figure*}

For the Fourier approach we focus on the domain $\bm{x} \in [0, 1]^2$ and choose $\bm{\mu} = (0.5, 0.5)^T$ and, $\Sigma = \begin{pmatrix} 0.05 & 0 \\ 0 & 0.05\end{pmatrix}$. Note that the bivariate Student's t-distribution does not factorize into two univariate Student's t-distributions. First, we periodically extend the function from $[0, 1]^2$ to $[-1, 1]^2$, interpolate it with the Fourier series in Eq.~\eqref{eq:f_d_Fourier_2d} and compute coefficients with the fast Fourier transform following Appendix~\ref{app:preprocessing}. We choose the same degree and same grid size in each variable, i.e. $d_x=d_y=d$ and $n_x=n_y=n$. Figure~\ref{fig:Cauchy-fourier-results} shows the results of approximating the periodically extended Student's t-distribution by a Fourier series of varying degree $d$ and for different grid sizes $2^n \times 2^n$. The maximum error in Fig.~\ref{fig:Cauchy-fourier-results}(b) scales asymptotically as $\mathcal{O}(1/d)$. This is caused by the periodic extension introducing a discontinuity in the first derivative of the extended function at the boundaries $x=0, 1$ and $y=0, 1$. In the univariate case a rule-of-thumb is that for $k-1$ times continuously differentiable functions and a $k$-th derivative with bounded variation, the asymptotic interpolation error is $\mathcal{O}(1/d^k)$~\cite[see Theorem 7.2 for an equivalent statement for Chebyshev interpolants]{trefethenApproximationTheoryApproximation2019}. Our numerics indicate this also holds for the bivariate case of the considered function. To reach higher convergence rate we could e.g. adjust the periodic extension method to smoothen the sharp corners at the boundaries.

Figure~\ref{fig:Cauchy-fourier-results}(c) shows the number of two-qubit gates after compilation to the Quantinuum native gate set with optimization level 2. A least-square fit of the data to the scaling of two-qubits gates derived in Sec.~\ref{sec:scaling}, $\alpha_0 + \alpha_1 D n \lceil \log_2(2d+1) \rceil + \alpha_2 2^{D\lceil \log_2(2d+1) \rceil}$, shows excellent agreement. The small coefficients $\alpha_1 < 1$ (see caption of Fig.~\ref{fig:Cauchy-fourier-results}) indicate savings from the compilation step. As in the Chebyshev case, $\alpha_2 \approx 3$ indicates worst-case synthesis of the operators $A$, $C$, $A^\dagger$ in the LCU on the coefficient registers. The largest circuit in Fig.~\ref{fig:Cauchy-fourier-results}(c) at $d=63$, $n=6$ requires 26 qubits -- 12 for the main register and 14 for the ancilla register, cf.  Table~\ref{tab:resources}. Similar to the Chebyshev case the success probability in Fig.~\ref{fig:Cauchy-fourier-results}(d) saturates close to its asymptotic value of $p_\text{success}^*=7.52\%$. This is computed as $p_\text{success}^* = \int_{[0,1]^2}\diff x\diff y |f(x, y)|^2 / \big(\sum_{k,l=-63}^{63} |c_{k,l}|\big)^2$. The sum runs to degree $63$ because the coefficients beyond this degree did not contribute noticeably (tested up to $k,l=128$) and fall off with increasing degree.

\subsection{Single electron in periodic 3D Coulomb potential}

Our protocols are applicable even when the exact target function is unknown, but we have access to the coefficients of an approximation by other means, e.g. through the solution of a differential equation. To demonstrate this, in this section we use our Fourier approach for the preparation of wavefunctions of a single electron in a 3D translationally invariant Coulomb potential.

The prepared quantum state can serve as an initial state for the quantum simulation of materials in a unit cell of atoms with nuclear electronic Coulomb interactions. The unit cell wavefunction with cell volume equal to 1, is expressed as a linear combination of $N^3$ plane wave basis functions via
\begin{equation}
\label{eqn:pw_chem}
    \psi(\bm{r}) = \sum_{\bm{k}\in G} c_{\bm{k}} \e^{\im \pi \bm{k}^T \bm{r}}.
\end{equation}
Here, $\bm{k} = (k_x, k_y, k_z)^T$ runs over $G = \left[ -\frac{N}{2}, \frac{N - 1}{2} \right]^3 \subset \mathbb{Z}^3$, $\bm{r}=(x, y, z)^T \in [0, 1]^3$ is the real-space position of the electron in the unit cell. Equation~\eqref{eqn:pw_chem} is a trigonometric polynomial in 3 variables with maximal degree $N/2$ in each dimension, to which we apply our Fourier approach.

In order to obtain the coefficients $c_{\bm{k}}$, we insert the expansion~\eqref{eqn:pw_chem} into the time-independent Schr\"odinger equation and solve the resulting set of linear equations. In the plane wave basis the kinetic energy matrix is diagonal with entries given by
\begin{equation}
    T_{\bm{k},\bm{k}} = \frac{(\hbar \pi  \| \bm{k}\|)^2}{2m} \dyad{\bm{k}}{\bm{k}}.
\end{equation}

The electron-nuclear attraction matrix elements have the following convenient analytical form in the basis of plane waves 
\begin{equation}
    U_{\bm{k},\bm{k}'} = \frac{4}{\pi} \sum_{\ell=1}^{L} w_{\ell} \frac{\e^{\im\pi(\bm{k}' - \bm{k})^T \mathbf{R}_{\ell}}}{\| \bm{k} - \bm{k}' \|^2} \dyad{\bm{k}}{\bm{k}'}.
\end{equation}
$\mathbf{R}_\ell$ is the position of nucleus $\ell$ in real space and $w_\ell$ is the corresponding nuclear weight and $L$ is the number of nuclei. The $\pi$ factors arise from the definition of the state in Eq.~\eqref{eqn:pw_chem}.

The electronic Hamiltonian matrix elements are then given by $H_{\bm{k},\bm{k}'}^\mathrm{(el)} = -T_{\bm{k},\bm{k}}\delta_{\bm{k},\bm{k}'}  - U_{\bm{k},\bm{k}'}$ and the time-independent Schr\"odinger equation becomes $H^\mathrm{(el)}\bm{c} = E\bm{c}$ with $\bm{c}$ the vector of coefficients in Eq.~\eqref{eqn:pw_chem}. Solutions $\bm{c}$ for the first and second lowest eigenenergies result in the upper bounds to the ground and excited state wavefunction of the non-interacting single electronic system. We use those coefficients in our multivariate Fourier approach to prepare the quantum states for the ground and first excited state wavefunctions.

\begin{figure}[t]
    \centering
    \includegraphics[width=\linewidth]{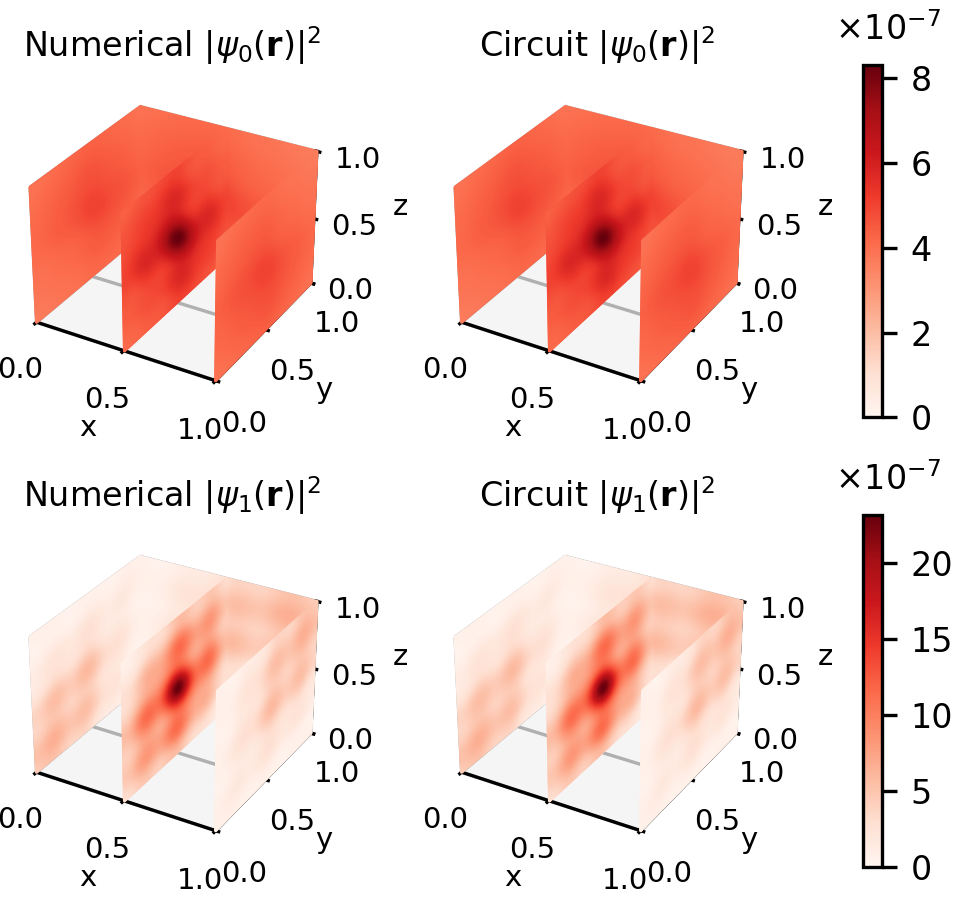}
    \caption{Probability density of the ground state (top panels) and first excited state (bottom) electronic wavefunctions for a unit cell with a single nucleus at the center at $(0.5,0.5,0.5)$. Left panels: direct evaluation of Eq.~\eqref{eqn:pw_chem} with a total of $N^3=512$ coefficients from the numerical solution of the time-independent Schr\"odinger equation. Right panels: equivalent solutions from the Fourier approach of our method evaluated with a statevector simulator using $n=7$ qubits per spatial dimension and $a=3$ ancilla qubits per dimension for a total of 30 qubits.}
    \label{fig:3Dunit}
\end{figure}

We choose the simplest 3D periodic unit cell for demonstration, which is a single Coulomb potential of nuclear weight and charge $1$ at the center of a $(1\times1\times1)$ unit cell at position $(0.5,0.5,0.5)$. Other more advanced unit cells can be used, such as face-centered cubic (FCC). In the left panels of Fig.~\ref{fig:3Dunit} we present the probability density $|\psi(\bm{r})|^2$ of the ground (top) and first excited state wavefunctions (bottom) computed directly with Eq.~\eqref{eqn:pw_chem} and coefficients from the solution of the time-independent Schr\"odinger equation above. We then use the same set of coefficients to construct the circuits of our Fourier-based protocol for preparing those wavefunctions.  In the right panels of Fig.~\ref{fig:3Dunit} we present the probability density of the resulting wavefunctions obtained from a statevector simulation of those circuits. The simulated circuits replicate to numerical precision the results from the direct numerical computation.

\begin{table*}[t]
    \centering
    \begin{tabular}{r|c|c|c|c|c|c}
    \toprule
    $n$ &
    $a$ &
    $D$ &
    Grid points ($2^{Dn}$) & 
    Coefficients ($2^{Da}$) & 
    Total qubits ($Dn + Da$) & 
    Two-qubit gates
    \\
    \midrule
    4 &
    3 &
    3 &
    \num{4096} & 
    \num{512} & 
    21 & 
    \num{1602}  
    \\
    5 &
    3 &
    3 &
    \num{32768} & 
    \num{512} & 
    24 & 
    \num{1620}  
    \\
    6 &
    3 &
    3 &
    \num{262144} & 
    \num{512} & 
    27 & 
    \num{1638}  
    \\
    * 7 &
    3 &
    3 &
    \num{2097152} & 
    \num{512} & 
    30 & 
    \num{1656}
    \\
    7 &
    4 &
    3 &
    \num{2097152} & 
    \num{4096} & 
    33 & 
    \num{12450}
    \\
    \bottomrule
    \end{tabular}
    \caption{Circuit compilation results for the solutions of the Schr{\"o}dinger equation in a 3D periodic Coulomb potential with varying numbers of grid points and Fourier coefficients. With $n$ the number of qubits per spatial dimension and $a=\lceil \log_2 N\rceil$ the number of ancillas per dimension the total number of qubits is $3n+3a$. The number of two-qubit gates is determined by compiling the circuits to the gate set given in the main text. The row marked with * corresponds to the setting of Fig.~\ref{fig:3Dunit}.}
    \label{tab:colombresults}
\end{table*}

Table~\ref{tab:colombresults} shows the resource requirements from compilation of the circuits of our method for preparing the wavefunctions with different numbers of grid points and coefficients. \texttt{pytket}~\cite{sivarajahVertKetRangle2020} is used to construct the circuits, using the gateset
\{\CX, \Rz, \Rx{}\}. This confirms that the dominant cost of the algorithm comes from the addition of the ancilla qubits that contain the Fourier coefficients, which is expected from Table~\ref{tab:resources}. Increasing the spatial resolution (given by $2^{3n}$, where $n$ is the number of qubits per spatial dimension), leads to little increase in gate count due to the efficient control structure presented in Fig.~\ref{fig:factorize_fourier}. Using a total of 9 and 12 ancilla qubits (i.e. registers of size $a=3, 4$ per dimension) allows one to encode \num{512} and \num{4096} Fourier coefficients, respectively. Thus, the large increase in the number of gates is due to the 9 vs. 12 qubit circuits for $A$, $C$ and $A^\dag$ (cf. the term $\mathcal{O}(d^D)$ for two-qubit gates in Table~\ref{tab:resources}). For the 30 qubit case marked with *, the success probability of generating the ground ($\psi_0$) and first excited ($\psi_1$) states are \num{0.7047} and \num{0.2392}, respectively.

Although these are only single electron wavefunctions, they can be used as a starting point for many-electron wavefunctions by anti-symmetrizing the spatial registers of multiple single electron states following the procedure proposed in Ref.~\cite{grid1q}, or as the initial state for real or imaginary time evolution under an interacting electron Hamiltonian in the first quantized picture. 

\begin{figure*}[t]
    \centering
    \includegraphics[width=\textwidth]{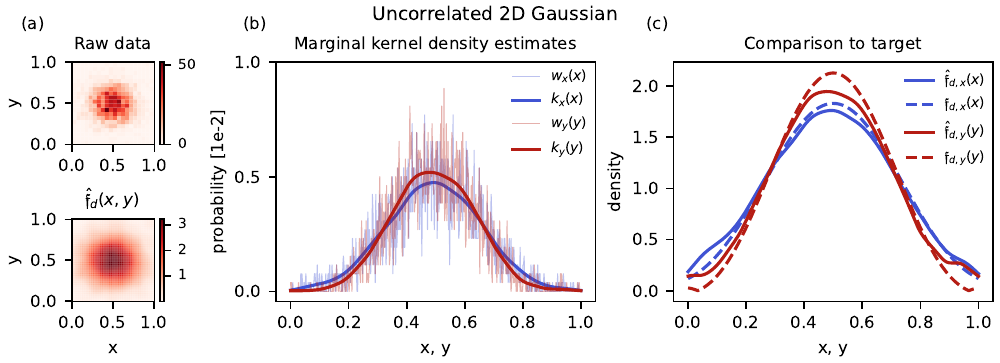}
    \includegraphics[width=\textwidth]{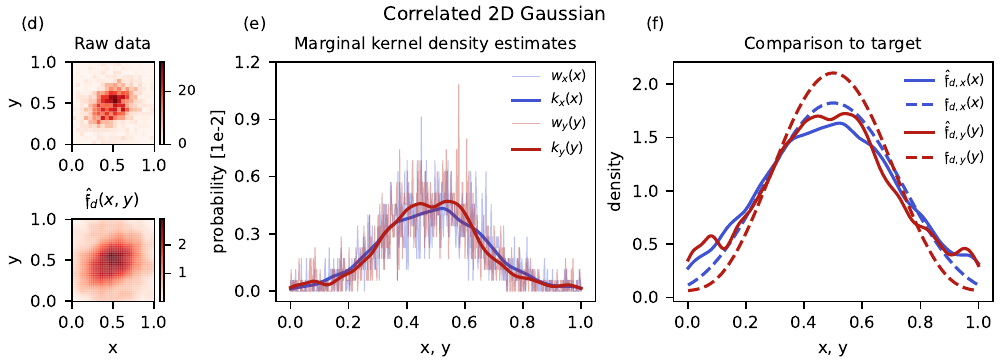}
    \caption{
    Experimental results for preparing a 2D Fourier series approximation of the bivariate Gaussian distribution. We show two setting, an uncorrelated Gaussian (upper panel) and a correlated Gaussian (lower panel); parameters are given in the main text.
    (a, d) Histograms of the raw two-dimensional measurement counts with 50 bins along each axis (upper plots), and the resulting estimates $\hat{\mathfrak{f}}_d(x,y)$ (lower plots, obtained from kernel density estimates) of the target functions $\mathfrak{f}_d(x,y)$.
    (b, e) Marginal probabilities $w_x(x), w_y(y)$ of the measurement data (thin solid lines) and the corresponding kernel density estimates $k_x(x), k_y(y)$ (bold solid lines).
    (c, f) Direct comparison between marginal estimates $\hat{\mathfrak{f}}_{d,x}(x), \hat{\mathfrak{f}}_{d,y}(y)$ (solid lines) and the target functions' marginals $\mathfrak{f}_{d,x}(x), \mathfrak{f}_{d,y}(y)$ (dashed lines).
    }
    \label{fig:hardware-results}
\end{figure*}

\begin{table*}[t]
    \centering
    \def\arraystretch{1.5}
    \begin{tabular}{cc|cccccccc}
    \toprule
    && $\mu_{d,x}, \hat{\mu}_{d,x}$ & $\mu_{d,y}, \hat{\mu}_{d,y}$ & $\sigma_{d,x}^2, \hat{\sigma}_{d,x}^2$ & $\sigma_{d,y}^2, \hat{\sigma}_{d,y}^2$ & $\rho_d, \hat{\rho}_d$ & $F_x$ & $F_y$ & $F$ \\
    \midrule
    \multirow[c]{2}{.8in}[0in]{Uncorrelated\\Gaussian} & \multicolumn{1}{|c|}{$\mathfrak{f}_d(x,y)$}
    & $0.5$ & $0.5$ 
    & $0.0413$ & $0.0299$ & $0.0$
    & \multirow[c]{2}{*}[0in]{$0.998$}
    & \multirow[c]{2}{*}[0in]{$0.984$}
    & \multirow[c]{2}{*}[0in]{$0.986$} \\
    & \multicolumn{1}{|c|}{$\hat{\mathfrak{f}}_d(x,y)$}
    & $0.494$ & $0.495$ 
    & $0.0468$ & $0.0396$
    & $-0.015$ &&& \\ \hline
    \multirow[c]{2}{.8in}[0in]{Correlated\\Gaussian} & \multicolumn{1}{|c|}{$\mathfrak{f}_d(x,y)$}
    & $0.5$ & $0.5$ 
    & $0.0415$ & $0.0321$ & $0.353$
    & \multirow[c]{2}{*}[0in]{$0.991$}
    & \multirow[c]{2}{*}[0in]{$0.963$}
    & \multirow[c]{2}{*}[0in]{$0.918$} \\
    & \multicolumn{1}{|c|}{$\hat{\mathfrak{f}}_d(x,y)$}
    & $0.493$ & $0.488$ 
    & $0.0519$ & $0.0510$
    & $0.088$ &&& \\
    \bottomrule
    \end{tabular}
    \caption{
    Experimental results for preparing a 2D Fourier series approximation of the bivariate Gaussian distribution, see Fig.~\ref{fig:hardware-results}. We consider two settings, the preparation of an uncorrelated and a correlated Gaussian.
    The table collects, for both settings, mean values, variances and correlation coefficients [cf. Eq.~\eqref{eq:statistical_quantifiers}] of the target densities $\mathfrak{f}_d(x,y)$ and their estimates $\hat{\mathfrak{f}}_d(x,y)$, based on kernel density estimation.
    Moreover, we show the fidelities $F_x, F_y$ [cf. Eq.~\eqref{eq:class_fidelity}] between marginal estimates, $\hat{\mathfrak{f}}_{d,x}, \hat{\mathfrak{f}}_{d,y}$, and target functions' marginals, $\mathfrak{f}_{d,x}, \mathfrak{f}_{d,y}$, and the full fidelity $F$ between $\hat{\mathfrak{f}}_d$ and $\mathfrak{f}_d$ [cf. Eq.~\eqref{eq:class_fidelity}].
    }
    \label{tab:hardware_experiment}
\end{table*}

\subsection{Bivariate Gaussian on quantum hardware}
\label{sec:hardware_demonstration}

This section demonstrates our state preparation protocol on the H2-1 trapped-ion quantum computer and illustrates the effect of noise. We choose to use the Fourier basis functions as they lead to circuits with fewer two-qubit gates. We focus on analyzing the output signal without any error mitigation. Error mitigation protocols, if needed, would likely be performed in combination with a downstream quantum algorithm, not after state preparation, which is only the first step of the computation.

We choose a two-dimensional Gaussian distribution as a target function. The low overheads of our method allow us to prepare the Gaussian on a fine grid with 24 qubits using up to \num{237} two-qubit gates with meaningful output signal.
The Gaussian distribution is given by
\begin{align}
\label{eq:hardware_demonstration_gaussian}
    f(\bm{x}) = \frac{1}{2 \pi \sqrt{\det(\Sigma)}} \exp[-\tfrac{1}{2}(\bm{x}-\bm{\mu})^T \Sigma^{-1}(\bm{x}-\bm{\mu})],
\end{align}
where $\bm{x} = (x,y)^T$, $\bm{\mu}$ is the mean and $\Sigma$ is the positive semi-definite covariance matrix.
Let $Z = X \times Y$ be a 2D random variable distributed according to $f(x,y)$.
The covariance matrix can be parametrized by the marginal variances $\sigma_x^2 = \mathrm{Var}(X)$ and $\sigma_y^2 = \mathrm{Var}(Y)$ (diagonal entries of $\Sigma$),
and the (Pearson) correlation coefficient $\rho = \mathrm{Cov}(X, Y) / \sigma_x \sigma_y$ (setting off-diagonal entries $\rho \sigma_x \sigma_y$).
For this demonstration, we use $\bm{\mu} = (\mu_x, \mu_y)^T = (0.5, 0.5)^T$, $\sigma_x = 0.22, \sigma_y = 0.18$, and the two choices $\rho = 0.0$ and $\rho = 0.4$.
The first setting defines an uncorrelated Gaussian, i.e. the bivariate density $f(x,y)$ is the product of two univariate densities.
The second setting defines a correlated Gaussian, which does not factorize.

For both settings we use $n_x = n_y = 9$ qubits for spatial discretization of the domain $[0,1]^2$ (cf. Fig.~\ref{fig:quantum_circuit_2d}) into $\num{512} \times \num{512} = \num{262144}$ grid points (for comparison, Ref.~\cite{Moosa2023} prepared a bivariate function on a grid of size $32 \times 32$).
We use a Fourier series approximation $f_d(x,y)$ of degrees $d = d_x = d_y = 3$, which leads to a total of $(2d + 1)^2 = 49$ coefficients.
Note that we use a different method to compute the Fourier coefficients than the one used in Sec.~\ref{ssec:student-t} because the approximation at the chosen degree is better for the specific target, see Appendix~\ref{app:experiment_implementation_details}.
To encode the Fourier coefficients we require ancilla registers with $a_x = a_y = 3$ qubits each, cf. Fig.~\ref{fig:quantum_circuit_2d}.
Overall, this results in a circuit with $24$ qubits.
We test our approach on the H2-1 trapped-ion quantum computer~\cite{Moses2023,H2datasheet}. At the time of this demonstration in April and May 2024, it provides $32$ fully-connected qubits with average single- and two-qubit gate infidelities of $\num{3e-5}$ and $\num{2e-3}$, respectively, and average state preparation and measurement error of $\num{2e-3}$. Detailed specifications are given in Appendix~\ref{app:experiment_specifications}.

The Fourier approximation $f_{d=3}(x,y)$ of the (positive) Gaussian density in Eq.~\eqref{eq:hardware_demonstration_gaussian} can have noticeable negative values in $[0,1]^2$.
This is visible in the small kinks of the red dashed line (showing a marginal over $|f_d(x,y)|$) in Fig.~\ref{fig:hardware-results}(c) close to the boundaries of the domain $[0,1]$.
To avoid expensive state-tomographic methods, we only read out the magnitudes $|f_d(x,y)|$ of the Fourier series via measurements in the computational basis and compare our experimental data to the positive target function $\mathfrak{f}_d(x,y) = |f_d(x,y)|$.
We run each experiment with $\num{20000}$ shots. The success probability in our setting is around $10\%$ (for details see Appendix~\ref{app:experiment_implementation_details}).
Let $w(x_i,y_j)$ be the probability to measure grid point $(x_i,y_j) \in [0,1]^2, i,j = 1, \dots, 512$, after post-selection on the zero state in the ancilla register (cf. Fig.~\ref{fig:quantum_circuit_1d}), and let $w_x(x_i) = \sum_{j=1}^{512} w(x_i, y_j)$, $w_y(y_j) = \sum_{i=1}^{512} w(x_i, y_j)$ be its marginals. The upper panels of Fig.~\ref{fig:hardware-results}(a) and (d) show histograms of the joint measurement distribution $w(x,y)$ for the uncorrelated and correlated settings, respectively. Figures~\ref{fig:hardware-results}(b) and (e) show the corresponding $x$- and $y$-marginals as thin lines.
Notably, the measured signals suffer from substantial fluctuations due to shot noise.

To compare the empirical probabilities to the target $\mathfrak{f}_d(x,y)$ in a meaningful way, we smoothen the experimental data with kernel density estimation (KDE) ~\cite{duin_choice_1976,rudemo_empirical_1982,hastie_elements_2009}.
The KDE yields a continuous approximation of the density underlying a finite data sample. This is obtained by replacing the raw counts with, in our case, Gaussian kernels of a given bandwidth $h$ (see Appendix~\ref{app:experiment_kde}).
The KDE becomes smoother with increasing bandwidth $h$. On the other hand, for too large $h$ the KDE loses information about the underlying data.
In the limit $h \rightarrow \infty$, the KDE approaches a perfectly flat distribution. The ``optimal'' bandwidth can be computed via minimization of an appropriate cost function.
A naive choice would be, for example, to consider a distance measure or statistical divergence (such as the relative entropy) between KDE and the target function $\mathfrak{f}_d(x,y)$. However, this choice introduces an obvious bias.
Instead we choose to optimize $h$ via cross validation from the measurement data alone (see Appendix~\ref{app:hardware_experiment}).

Let $k(x,y)$, $k_x(x)$ and $k_y(y)$ be the KDE [cf. Eq.~\eqref{eq:app_kde_estimate}] corresponding to the empirical probabilities $w(x_i, y_j)$, $w_x(x_i)$ and $w_y(y_j)$ ($i,j = 1,\dots, 512$), rescaled such that $\sum_{i,j=1}^{512} k(x_i, y_j) = \sum_{i=1}^{512} k_x(x_i) = \sum_{j=1}^{512} k_y(y_j) = 1$.
The thick solid lines in Figs.~\ref{fig:hardware-results}(b) and~(e) are the marginal KDE $k_x(x_i)$ and $k_y(y_j)$.
To compare to $\mathfrak{f}_d(x,y)$, we need to take the square root, which defines our experimental estimates $\hat{\mathfrak{f}}_d(x , y) = \sqrt{k(x, y)}$, $\hat{\mathfrak{f}}_{d,x}(x) = \sqrt{k_x(x)}$ and $\hat{\mathfrak{f}}_{d,y}(y) = \sqrt{k_y(y)}$.
The lower panels of Fig.~\ref{fig:hardware-results}(a) and ~(d) display $\hat{\mathfrak{f}}_d(x,y)$.
In Fig.~\ref{fig:hardware-results}(c) and ~(f), we compare the estimated marginals (solid lines) to the target marginals $\mathfrak{f}_{d,x}, \mathfrak{f}_{d,y}$ of $\mathfrak{f}_d$ (dashed lines).
For direct comparison, we rescale all functions $\hat{\mathfrak{f}}_d, \hat{\mathfrak{f}}_{d,x}, \hat{\mathfrak{f}}_{d,y}, \mathfrak{f}_{d,x}, \mathfrak{f}_{d,y}$ such that they integrate to one on their corresponding domains, i.e. that they represent proper probability densities.
Visually, we observe a good agreement.
We clearly recognize an elongated shape in the lower plot of Fig.~\ref{fig:hardware-results}(d), as expected from a correlated 2D Gaussian.

For a quantitative comparison we determine the mean values and the covariance matrix of the target function $\mathfrak{f}_d(x,y)$ and its estimate $\hat{\mathfrak{f}}_d(x,y)$ (obtained from the experimental data):
\begin{equation}
\begin{split}
\label{eq:statistical_quantifiers}
    \mu_{d,x} &= \int_{[0,1]^2}\diff x\diff y\, x \mathfrak{f}_d(x,y) , \\
    \mu_{d,y} &= \int_{[0,1]^2}\diff x\diff y\, y \mathfrak{f}_d(x,y) , \\
    \sigma_{d,x}^2 &= \int_{[0,1]^2}\diff x\diff y\, (x - \mu_{d,x})^2 \mathfrak{f}_d(x,y) , \\
    \sigma_{d,y}^2 &= \int_{[0,1]^2}\diff x\diff y\, (y - \mu_{d,y})^2 \mathfrak{f}_d(x,y) , \\
    \rho_d &= \frac{\int_{[0,1]^2}\diff x\diff y\, (x - \mu_{d,x})(y - \mu_{d,y}) \mathfrak{f}_d(x,y)}{\sigma_{d,x} \sigma_{d,y}}  ,
\end{split}
\end{equation}
and $\hat{\mu}_{d,x}, \hat{\mu}_{d,y}, \hat{\sigma}_{d,x}^2, \hat{\sigma}_{d,x}^2, \hat{\rho}_d$ computed equivalently by replacing all quantities with their ``hat'' versions (e.g. replace $\mathfrak{f}_d(x,y)$ with $\hat{\mathfrak{f}}_d(x,y)$).
Furthermore, we compute the classical fidelity between the target and measurements
\begin{equation}
\label{eq:class_fidelity}
    F = \frac{\left( \sum_{i,j = 1}^{512} \sqrt{\mathfrak{f}_d(x_i, y_j) \hat{\mathfrak{f}}_d(x_i, y_j)} \right)^2}{\sum_{i,j=1}^{512} \mathfrak{f}_d(x_i, y_j) \sum_{i,j = 1}^{512} \hat{\mathfrak{f}}_d(x_i, y_j)},
\end{equation}
where $x_i, y_i$ runs over the grid points.
Table~\ref{tab:hardware_experiment} compares the results for both settings.
Overall, we observe a good agreement between the targets and the experimentally prepared functions, both in terms of statistical quantifiers and fidelities.
For the uncorrelated 2D Gaussian we obtain an overall fidelity of $F = 0.986$, for the correlated one we obtain $F = 0.918$.

We observe a larger deviation between target and prepared function for the correlated Gaussian.
For example, the observed variances $\hat{\sigma}_{d,x}^2, \hat{\sigma}_{d,x}^2$ overestimate the target variances $\sigma_{d,x}^2, \sigma_{d,x}^2$ in this setting.
Although we clearly identify a non-zero correlation coefficient in the setting of the correlated Gaussian, the estimated $\hat{\rho}_d = 0.088$ is significantly smaller, roughly by a factor of four, than the target value $\rho_d = 0.353$.
The pronounced discrepancy between $\mathfrak{f}_d$ and $\hat{\mathfrak{f}}_d$ in the correlated setting, in comparison to the uncorrelated Gaussian, is expected.
This is because, in the latter scenario, we are able to exploit the factorization of $\mathfrak{f}_d(x,y)$ into two univariate functions also at the level of circuit synthesis. Indeed we obtain much shallower circuits for the uncorrelated setting, which are much less prone to noise, as discussed in more detail in Appendix~\ref{app:experiment_implementation_details}.

\section{Conclusions}

This work introduces methods for the preparation of quantum states that encode multivariate functions, and analyzes their resource requirements.
As a proof-of-concept, we demonstrate the successful preparation of representative initial states (with potential applications e.g. in quantum chemistry, physics and finance simulations) on comparatively high-resolution grids.
We execute a 24-qubit experiment on Quantinuum's H2-1 trapped-ion quantum computer and analyse the impact of noise on the prepared states. The code implementing the presented protocols and the data to reproduce all results are available at GitHub and Zenodo~\cite{mvsp1.0.0.2025}.

Many functions of interest cannot be naturally encoded in a quantum state (e.g. $1/x$ due to a singularity), and function approximation becomes a necessary step to obtain practical protocols. Our methods are based on multivariate versions of Fourier and Chebyshev series approximations. We note that power series can be mapped to Chebyshev series~\cite{Thacher_1964,Nyengeri_2021} where our method applies as-is. Walsh series used e.g. in~\cite{zylberman2023efficient,Welch_2014} can be implemented as a special case of our Fourier method. It is also possible to use different basis functions for each dimension in a hybrid Fourier-Chebyshev method. Moreover, the building blocks of our method can be applied recursively to create more expressive basis functions, e.g. wavelets and kernels, which are then linearly combined to form more complex functions. Ultimately, the approximation method and error budget will depend on the use case (e.g. whether the target function is periodic) and specifics of the available quantum hardware (e.g. two-qubit gate fidelity). 

QSP~\cite{Low_2017,Gilyen_2019,Silva2021,Martyn2021,Kikuchi_2023} has attracted significant interest as a framework for developing quantum algorithms via polynomial transformations of block-encoded matrices. Multivariate versions of QSP have been proposed and analyzed, although questions about expressive power remain open~\cite{Rossi2022, mori2023comment, nemeth2023variants}. Our work shows that linear combination of unitaries~\cite{Childs2012}, with its simplicity and effectiveness, is a valid alternative for the task of preparing quantum states that encode multivariate functions in their amplitudes. An interesting research direction consists of using (univariate) QSP to further process the matrices produced by our method. For example, suppose we wanted to prepare a state that encodes a hypersurface -- a set of dimension $(D-1)$ defined by a polynomial equation of the form $f(x_1, \dots, x_D) = 0$. After discretizing each dimension via $H^{\otimes D}$ and using our method to construct the diagonal operator $f(H^{\otimes D})$, we use QSP to filter its eigenvalues. In particular, we want a polynomial $p$ such that $p(0)=1$ and $|p(y)|$ is small for $y \in [-1, -\epsilon] \cup [\epsilon, 1]$. An optimal filter is given by Lin and Tong in~\cite{Lin2020optimalpolynomial}. Applying the resulting operator to the uniform superposition we obtain a state proportional to $\sum_{x_1} \cdots \sum_{x_D} p( f(x_1, \dots, x_D) ) \ket{x_1, \dots, x_D}$, approximately encoding the hypersurface. Remarkably, this proposal uses univariate QSP for a fundamentally multivariate problem. 

\section*{Acknowledgements}
We thank Hitomi Mori for useful discussions, and Alexandre Krajenbrink and Enrico Rinaldi for helpful feedback on the manuscript.

\bibliographystyle{quantum}
\bibliography{main}

\onecolumn
\appendix

\section{Classical preprocessing for the Chebyshev method}
\label{app:Chebyshev}

\subsection{One dimension}
A Lipshitz continuous function $f$ defined on $[-1, 1]$ has a unique expansion as a Chebyshev series,
\begin{equation}
    f(x) = \sum_{k=0}^\infty \hat c_k T_k(x)
\end{equation}
with coefficients
\begin{align}\label{eq:cheb_coeff}
    \hat c_k &= \frac{2-\delta_{k,0}}{\pi} \int_{-1}^1 \frac{\diff x}{\sqrt{1 - x^2}}f(x) T_k(x),
\end{align}
where $\delta_{k,j}$ is the Kronecker delta (see e.g. \cite[Theorem 3.1]{trefethenApproximationTheoryApproximation2019}).
The Chebyshev polynomials $T_k(x)$ satisfy the orthogonality condition
\begin{align}
    \int_{-1}^{1}\frac{\diff x}{\sqrt{1-x^2}}T_k(x)T_j(x) = 
    \begin{cases}
        0 & \text{if } k\neq j,\\
        \frac{\pi}{2} & \text{if } k=j\neq 0,\\
        \pi & \text{if } k=j=0
    \end{cases}
\end{align}
as well as the discrete orthogonality condition
\begin{equation}
    \sum_{m=0}^{N-1} T_k\left(x_m^{(N)}\right) T_j\left(x_m^{(N)}\right) = \begin{cases}
        0 & \text{if } k\neq j,\\
        \frac{N}{2} & \text{if } k=j \neq 0,\\
        N & \text{if } k=j=0,
    \end{cases}
\end{equation}
where $x_m^{(N)} = \cos\big(\frac{\pi(2m+1)}{2N} \big)$ for $m=0, 1, \dots, N-1$ are the roots of the Chebyshev polynomial $T_N(x)$.

The analysis can be extended to complex functions. The Chebyshev series remains convergent within an ellipse in the complex plane with foci $\pm 1$ that does not contain any poles, branch points or other singularities of $f$~\cite[Theorem 7]{boydChebyshevFourierSpectral2001}.

For the numerical results in this paper we construct the maximal degree-$d$ polynomial approximation Eq.~\eqref{eq:f_d_Chebyshev} via interpolation.
The interpolant $f_d$ is uniquely determined by finding coefficients $c_k$ such that $f_d(x_m) = f(x_m)$ on a set of $d+1$ points $\left\{x_m\right\}_{m=0}^d$. We choose the roots of the degree-$(d+1)$ Chebyshev polynomial as interpolation points, $\left\{x_m^{(d+1)}\right\}_{m=0}^d$, as this entails favorable convergence properties to the target function.

First, let us consider the interpolation error~\cite{Elliott1987}.
If the Chebyshev series of a function $f$ exists, the error is bounded by $\max_{x\in[-1,1]} |f(x) - f_d(x)| \leq 2\sum_{k=d+1}^\infty |\hat c_k|$ \cite[Theorem 21]{boydChebyshevFourierSpectral2001}. Typically, the magnitude of coefficients decreases rapidly in $k$ with a rate depending on the smoothness of $f$. Given a target function $f\in C^{(d+1)}[-1,1]$ (i.e. $f$ is a complex function whose $(d+1)$-st derivative exists and is continuous on the interval $[-1,1]$), the degree-$d$ Chebyshev interpolation incurs an error
\begin{align}
    \varepsilon_d := \max_{x\in[-1,1]}|f(x)-f_d(x)|
    = \frac{|f^{(d+1)}(\xi)|}{2^d(d+1)!}
\end{align}
for some $\xi\in (-1,1)$. The same formula holds for truncation instead of interpolation, possibly for some other $\xi$~\cite{Elliott1987}. If $M$ is an upper bound for $f^{(d+1)}$ on $[-1,1]$ (this exists because $f^{(d+1)}$ is continuous on the bounded interval $[-1,1]$), then the truncation and interpolant error is upper-bounded by
\begin{equation}
    \varepsilon_d \leq \frac{M}{2^d(d+1)!}.
\end{equation}
The take-away message is that the Chebyshev interpolant rapidly converges to the target function if it is sufficiently smooth. The function does not need to be periodic as is the case for the Fourier expansion.

The coefficients $c_k$ of the interpolant can be computed with the fast Fourier transform. The interpolant $f_d$ is a Chebyshev series for a polynomial of degree at most $d$. Its coefficients are given exactly by applying the $(d+1)$-point trapezoidal rule to the integral in Eq.~\eqref{eq:cheb_coeff} evaluated at the $d+1$ Chebyshev roots~\cite[Theorem 21] {boydChebyshevFourierSpectral2001} 
\begin{equation}
    c_k' = \frac{2}{d+1} \sum_{m=0}^d f_d\left(x_m^{(d+1)}\right) T_k\left(x_m^{(d+1)}\right) = \frac{2}{d+1} \sum_{m=0}^d f_d\left(x_m^{(d+1)}\right) \cos\left(k\pi\frac{2m+1}{2(d+1)} \right).
\end{equation}
This is the discrete cosine transform, which we evaluate via the fast Fourier transform with complexity $\mathcal{O}(d\log d)$. The notation $c_k'$ means we need to divide the $k=0$ term by $2$, $c_0=c_0'/2$ while $c_k=c_k'$ for $k>0$.  The coefficients of the interpolant and infinite Chebyshev series are related via a simple formula~\cite[Theorem 21]{boydChebyshevFourierSpectral2001}.

\subsection{Two and higher dimensions}

To compute the 2D interpolant $f_{d_x, d_y}(x,y)$, Eq.~\eqref{eq:f_d_Chebyshev_2d}, we match the target function at the grid points $\left\{\left(x_m^{(d_x+1)}, y_n^{(d_y+1)}\right)\right\}_{m=0,n=0}^{d_x, d_y}$ given by the Chebyshev roots of degrees $(d_x+1)$ and $(d_y+1)$. In analogy to the 1D case, the coefficients $c_{k,l}$ of this interpolant are given by the 2D cosine transform on this grid
\begin{equation}
    c_{k,l}' = \frac{4}{(d_x+1)(d_y+1)} \sum_{m=0}^{d_x}\sum_{n=0}^{d_y} f(x_m^{(d_x+1)}, y_n^{(d_y+1)}) \cos\left(k\pi\frac{2m+1}{2(d_x+1)} \right) \cos\left(l\pi\frac{2n+1}{2(d_y+1)} \right).
\end{equation}
This can be evaluated with the fast Fourier transform with cost $\mathcal{O}(d_x d_y \log(d_x d_y))$. The notation $c_{k,l}'$ means we need to divide the $k=0, l=0$ terms appropriately, $c_{k,l} = c_{k,l}'/(1+\delta_{k,0}) (1+\delta_{l,0})$.

The above procedure readily generalizes to higher dimensions. However, for high degrees or high dimensions it may become computationally more efficient to compute a low-rank approximation of $f$ and approximate the univariate, constituent functions as Chebyshev polynomials~\cite{townsendExtensionChebfunTwo2013}.

Error bounds are more difficult to establish rigorously for the multivariate case. Ref.~\cite{trefethenMultivariatePolynomialApproximation2017} shows that multivariate functions with a certain property display similar behavior of the maximum error as the univariate case, namely exponential convergence to the target function with increasing (total or maximal) degree.

\section{Classical preprocessing for the Fourier method}
\label{app:preprocessing}

Here we discuss two preprocessing steps that may be required for implementing the Fourier approach. We used those methods in Sec.~\ref{ssec:student-t}. First, we describe one way to periodically extend functions which are not periodic with period 2. Second, we show how to compute coefficients in a degree-$d$ Fourier series approximation via the fast Fourier transform.

Without loss of generality consider a \emph{bivariate} target function $f: [0, 1]^2 \rightarrow \mathbb{C}$, which is not periodic with period 2. To obtain a converging Fourier series, we turn the function into a periodic one. We begin by extending the domain to $[-1, 1]^2$ by considering the function $\tilde{f}(x,y) = f(|x|, |y|)$.
This step guarantees that the function is mirrored with respect to both coordinate axes and for both the real and imaginary part. There are no jump discontinuities in the resulting function, although the first derivative may be discontinuous. Next, we extend the domain to $\mathbb{R}^2$ by considering the function
\begin{align}
    \tilde{f}(x + 2n, y + 2m) =  \tilde{f}(x,y) \qquad \forall n,m \in \mathbb{N} ,
\end{align}
which is a periodic function of period $2$ in both variables. Then we use this extended target function for finding Fourier series coefficients.

Next we discuss a method for computing the coefficients of a degree-$d$ trigonometric polynomial approximation to a function via interpolation and the fast Fourier transform. For simplicity let us consider the one-dimensional case, i.e. a periodic function $f(x)$ with period 2 on the interval $[-1, 1]$ (after applying a periodic extension if necessary). We assume $f$ admits a Fourier series $f(x) = \sum_{k=-\infty}^\infty \hat c_k \e^{\im\pi kx}$. Its coefficients are given by
\begin{equation}
\label{eq:app_1D_Fourier_trafo}
    \hat c_k = \frac{1}{2} \int_{-1}^{1}\diff x f(x) \e^{-\im\pi kx}.
\end{equation}
If those coefficients are known, a valid approximation is truncation of the infinite Fourier series at degree $d$. If they are unknown, it is often numerically favourable to seek an interpolant $f_d$ of the form Eq.~\eqref{eq:f_d_Fourier} such that $f_d(x_m) = f(x_m)$ on the set of points $x_m = 2m/(2d+1)$ for $m=0, \dots, 2d$. The coefficients of $f_d$ are given by
\begin{equation}
\label{eq:f_d_coeffs}
    c_k = \frac{1}{2d+1} \sum_{m=0}^{2d} f(x_m) \e^{-\im \tfrac{2\pi k m}{2d+1}}.
\end{equation}
This follows from multiplying $f_d(x_m)$ by $\e^{-\im\pi k x_m}$, summing over $m$ and using discrete orthogonality of the Fourier basis functions:
\begin{align}
    \sum_{m=0}^{2d} f_d(x_m) \e^{-\im\pi k x_m} = \sum_{m=0}^{2d} \sum_{l=-d}^d c_l \e^{\im\pi x_m (l - k)} = \sum_{l=-d}^d c_l (2d+1) \delta_{l,k} = (2d+1) c_k.
\end{align}
Inserting $x_m$ yields Eq.~\eqref{eq:f_d_coeffs}. 

Note that Eq.~\eqref{eq:f_d_coeffs} is the discrete Fourier transform of $\{f(x_m)\}_{m=0}^{2d}$, which can be evaluated with cost $\mathcal{O}(d\log d)$ via the fast Fourier transform. This method generalizes to $D$ dimensions and use of the $D$-dimensional fast Fourier transform with cost $\mathcal{O}[d_1 d_2 \cdots d_D \log(d_1 d_2 \cdots d_D)]$.

We do not discuss the convergence of multivariate Fourier series in general. We shall mention however that the examples presented in this paper do enjoy this property. For bivariate functions in $L^p$, with $p>1$, and with a careful definition of the partial sums, the Fourier series converges as shown by Fefferman~\cite{Fefferman1971convergence,Fefferman1971divergence}. 

\section{Hardware experiment}
\label{app:hardware_experiment}

\subsection{Hardware specifications}
\label{app:experiment_specifications}

For the experimental demonstration of our state preparation algorithm, we use the H2-1 trapped-ion quantum computer.
Here we summarize the relevant specifications of H2-1 at the time of the experiments performed in April and May 2024. For details, see Ref.~\cite{H2datasheet}. H2-1 operates with 32 qubits implemented with $S_{1/2}$ hyperfine clock states of $^{171}\mathrm{Yb}^+$ ions.
The native gate set consists of single-qubit rotation gates $\PhasedX(\alpha, \beta) = \Rz(\beta) \Rx(\alpha) \Rz(-\beta)$ and two-qubit gates $\RZZ(\theta)=\e^{-\im\frac{\theta}{2}Z\otimes Z}$, parametrized by real angles $\alpha, \beta$ and $\theta$.
All the gate operations as well as initializations and measurements are performed in one of four gate zones, each operating in parallel. 
The native two-qubit gates can be applied to an arbitrary pair of qubits by shuttling ions to one of the gate zones, enabling all-to-all connectivity.
Average single- and two-qubit gate infidelities are typically $0.003\%$ and $0.2\%$, respectively. State preparation and measurement error is typically $0.2\%$ on average.

\subsection{Implementation details}
\label{app:experiment_implementation_details}

As discussed in Sec.~\ref{sec:hardware_demonstration}, we consider the preparation of an uncorrelated and a correlated 2D Gaussian distribution $f(x,y)$, Eq.~\eqref{eq:hardware_demonstration_gaussian}, with the same mean value $\bm{\mu} = (\mu_x, \mu_y)^T$, covariance matrix $\Sigma$ parameterized by the same marginal variances $\sigma_x^2$, $\sigma_y^2$ and two values of the correlation coefficient $\rho$.
We prepare both functions on a 2D grid with $512 \times 512$ grid points, requirering $n_x = n_y = 9$ qubits for the spatial discretization of both axes.

For both settings, we compute the Fourier series coefficients in $f_d(x,y)$ with maximal degree $d=d_x = d_y = 3$ using the following method. First, we note that both Gaussians are close to $0$ at the boundary of the domain $[0, 1]^2$. We, thus, extend $f(x,y)$ to $[-1, 1]^2$ by setting the function to $0$ in the extended domain. Next, assuming this function admits a Fourier series, its coefficients can be approximated via
\begin{align}
\label{eq:app_2D_Fourier_trafo}
    \hat{c}_{k,\ell} = \frac{1}{4}
    \int_{[-1,1]^2}\diff x\diff y\, f(x, y) \e^{-\im \pi (xk + y\ell)} \approx  \frac{1}{4}
    \int \diff x\diff y\, f(x, y) \e^{\im (x, y)^T \bm{\omega}} =
    \frac{1}{4} \e^{ \im \bm{\mu}^T \bm{\omega} - \frac{1}{2} \bm{\omega}^T \bm{\Sigma} \bm{\omega}} \,.
\end{align}
The first equality is a generalisation of Eq.~\eqref{eq:app_1D_Fourier_trafo} to the bivariate case. The second approximate equality follows from extending the integration domain to the full space $\mathbb{R}^2$, and setting $\bm{\omega} = -\pi(k, \ell)^T$. This integral is the characterisic function of the bivariate Gaussian, for which there exists an analytic formula, given by the last expression in Eq.~\eqref{eq:app_2D_Fourier_trafo}.
The coefficients go to $0$ exponentially in $|k|$ and $|l|$, allowing us to truncate the series at relatively small degrees $d_x$ and $d_y$. We numerically observe that this method leads to a faster convergence for our specific target function compared to the method used in Sec.~\ref{ssec:student-t}. This allows us to use a lower degree at a similar approximation quality. A lower degree reduces circuit depth which is crucial for the experimental implementation.

Note that in the setting of the uncorrelated Gaussian, the density $f(x,y)$ [cf. Eq.~\eqref{eq:hardware_demonstration_gaussian}] and its Fourier series approximation $f_d(x,y)$ factorize into a product of univariate Gaussian densities.
This factorization can be exploited in the circuit construction as well. More precisely, we  prepare the two univariate functions with two separate circuits as given in Fig.~\ref{fig:quantum_circuit_1d} (each with $a=3, n=9$ qubits for the ancilla and main registers, respectively) and concatenate those. 
This considerably simplifies the circuit and reduces the gate complexity, in comparison to the full 2D state preparation circuit, cf. Fig.~\ref{fig:quantum_circuit_2d}. For the uncorrelated Gaussian we obtain $123$ single-qubit gates ($\PhasedX$), $80$ two-qubit gates ($\RZZ$) and a depth of $42$ for the full state preparation circuit. For the correlated Gaussian we obtain $266$ single-qubit gates, $237$ two-qubit gates and circuit depth $352$.

The success probability of the algorithm is predominantly controlled by the weight $\sum_x \sum_y |f_d(x,y)|^2$ of the prepared function in the domain $[0,1]^2$, see Eq.~\eqref{eq:success-analytical}.
Based on a noiseless simulation of the circuit, the ideal success probabilities are given by $p_\text{success} = 0.134$ and $p_\text{success} = 0.139$ for the case of the correlated and uncorrelated Gaussian, respectively.
The interplay between noise and circuit depth also affects the success probability. Deeper circuits increase overall noise when executed on noisy hardware, which, in general, reduces the overlap of the prepared (mixed) state with the target state $\ket{0}^{\otimes (a_x + a_y)}$ in the ancilla registers, cf. Fig.~\ref{fig:quantum_circuit_2d}. Generally, this reduces the success probability.
Experimentally, we obtain a success probability $p_\text{success} = \num{0.1299}$ for the uncorrelated setting and a success probability $p_\text{success} = \num{0.0877}$ for the correlated setting.

\subsection{Kernel density estimation}
\label{app:experiment_kde}

Let $\bm{X} = \lbrace \bm{x}_1, \dots, \bm{x}_{N_\mathrm{data}} \rbrace,\, \bm{x}_i \in \mathbb{R}^D$ be the observed data points in a given $D$-dimensional domain.
Assume the data is independently drawn from a density $f(\bm{x})$.
The kernel density estimate (KDE) of $f$ based on dataset $\bm{X}$ is given by
\begin{equation}
\label{eq:kde_formula}
    k^{\bm{X}}_h (\bm{x}) = \frac{1}{N_\mathrm{data}} \sum_{i=1}^{N_\mathrm{data}} K \left( \frac{\bm{x} - \bm{x}_i}{h} \right) \,,
\end{equation}
where $K$ is the kernel, i.e. a smooth normalized function (which integrates to one over the given domain), which is often chosen to be Gaussian.
The parameter $h$ is the bandwidth.
The KDE yields a smooth approximation of the density $f$ underlying the observations $\bm{x}_i$.
The bandwidth $h$ needs to be chosen based on the competition between the smoothness of the estimate and the information it contains about the dataset.
For a small bandwidth $h$ the KDE tends to overfit the data, and in the limit of $h\rightarrow 0$ the KDE just represents the original dataset without smoothing.
For $h\rightarrow \infty$, on the other hand, the KDE approaches a flat distribution, which does not contain any information about the dataset $\bm{X}$.

We optimize $h$ from the data, i.e. we do not assume access to the underlying density $f$, using cross validation.
For this let $\bm{X}_i$ be the dataset obtained from $\bm{X}$ by excluding the point $\bm{x}_i$, and $k^{\bm{X}_i}_h$ the corresponding KDE.
The probability to observe the excluded point is given by $k^{\bm{X}_i}_h(\bm{x}_i)$.
We want to maximize this probability, averaged over all choices of $i$.
In practice one often works with log-probabilities. Note that, since the logarithm is a monotonically increasing function, maximizing $k^{\bm{X}_i}_h(\bm{x}_i)$ is equivalent to maximizing $\log k^{\bm{X}_i}_h(\bm{x}_i)$. This defines our final optimizer, the mean of log-probabilities taken over $i$:
\begin{equation}
\label{eq:kde_optimizer}
    q(h) = \frac{1}{N_\mathrm{data}} \sum_{i = 1}^{N_\mathrm{data}} \log k^{\bm{X}_i}_h(\bm{x}_i) \,.
\end{equation}
The optimal $\hat{h}$ is obtained by maximizing $q$ over $h$. The KDE estimate of the target density $f$ is then given by
\begin{equation}
\label{eq:app_kde_estimate}
    k^{\bm{X}} = k^{\bm{X}}_{\hat{h}} \,.
\end{equation}
For a given $h$, the log-probability $\log k^{\bm{X}_i}_h (\bm{x}_i)$ is an unbiased estimator (under the distribution of $\bm{X}$) of the Kullback-Leibler divergence (up to a sign and a constant additive factor depending on $f$) between $f$ and $k^{\bm{X}}_h$.
What this means is that $\mathbb{E}[ \log k^{\bm{X}_i}_h (\bm{x}_i)]$ asymptotically approaches $\mathbb{E}[ \int\diff\bm{x} f(\bm{x}) \log k^{\bm{X}}_h (\bm{x})]$ for $N_\mathrm{data} \rightarrow \infty$, where the expectation $\mathbb{E}[\cdot]$ is taken with respect to the distribution of $\bm{X}$, see e.g. \cite{duin_choice_1976,rudemo_empirical_1982}.
There exist alternatives to the log-probabilities which are, for example, more closely related to the mean integrated squared error (MISE) between the target density $f$ and the estimator \cite{rudemo_empirical_1982}, see also \cite{bowman_alternative_1984} for a comparison of both.

\begin{figure}[t]
    \centering
    \includegraphics[width=\textwidth]{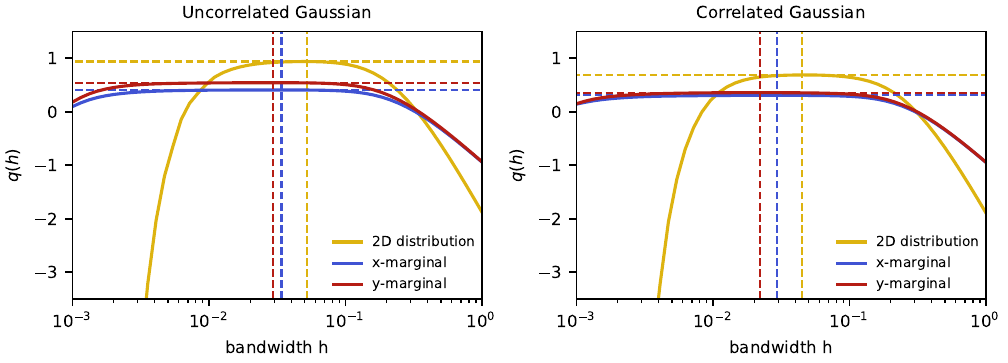}
    \caption{
    Cross validation of the kernel density estimates for $\mathfrak{f}_d(x,y) = |f_d(x,y)|$ and its marginals.
    The results are shown for the uncorrelated and the correlated Gaussian in the left and right plots, respectively.
    The maximal value of the mean of log-probabilities $q(h)$, Eq.~\eqref{eq:kde_optimizer}, and the corresponding bandwidth $h$ are highlighted via vertical and horizontal dashed lines. For the uncorrelated Gaussian, the maximum of $q(h)$ is achieved for $\hat{h} \approx 0.052$ for the full 2D distribution, and for $\hat{h}_x \approx 0.034$ and $\hat{h}_y \approx 0.029$ for its marginals.
    For the correlated Gaussian, we obtain optimal bandwidths $\hat{h} \approx 0.045$, $\hat{h}_x \approx 0.029$ and $\hat{h}_y \approx 0.022$.
    }
    \label{fig:kde_bandwidth_optimization}
\end{figure}

In Fig.~\ref{fig:kde_bandwidth_optimization} we apply KDE to the data obtained from our state preparation hardware experiments, cf. Sec.~\ref{app:experiment_implementation_details}.
To optimize $q(h)$ [cf. Eq.~\eqref{eq:kde_optimizer}] we use the \texttt{scikit-learn}~\cite{scikit-learn} implementations  of KDE and cross validation.
We first consider the datasets of the full 2D state preparation circuit. The dataset $\bm{X}$ is given by the two-dimensional points $\bm{x}_i = (x_i, y_i)$ (as represented via bitstrings, see Sec.~\ref{sec:discretization}) measured at the output of our state preparation algorithm, see Appendix~\ref{app:experiment_specifications} and \ref{app:experiment_implementation_details}.
In case of the uncorrelated Gaussian, we have $N_\mathrm{data} = \num{2598}$ data points, for the correlated one we have $N_\mathrm{data} = \num{1754}$.
The yellow solid curves in both plots of Fig.~\ref{fig:kde_bandwidth_optimization} show $q$ as a function of the bandwidth $h$, for both settings. We obtain optimal bandwidths (vertical yellow dashed lines) of $\hat{h} \approx 0.052$ (uncorrelated Gaussian) and $\hat{h} \approx 0.045$ (correlated Gaussian).
We repeat the same procedure for the marginals. The blue and red solid curves in Fig.~\ref{fig:kde_bandwidth_optimization} show $q(h)$ when we use $\bm{X} = \lbrace x_1, \dots, x_{N_\mathrm{data}} \rbrace$ (the x-marginal) and $\bm{X} = \lbrace y_1, \dots, y_{N_\mathrm{data}} \rbrace$ (the y-marginal), respectively.
Optimal marginal KDE are obtained for bandwidths $\hat{h}_x \approx 0.034, \hat{h}_y \approx 0.029$ for the setting of the uncorrelated Gaussian, and $\hat{h}_x \approx 0.029, \hat{h}_y \approx 0.022$ for the correlated Gaussian, as indicated by the vertical dashed blue and red lines in the plots, respectively.

\section{Comparison to prior work}
\label{app:prior_works}

Here we review prior works that (i) are not primarily heuristic and (ii) focus
on or can be extended in a straightforward way to multivariate functions.
Table~\ref{tab:prev_work} summarizes their resource estimates discussed in the
following subsections.

We consider a multivariate function $f$ defined on $[-1,1]^D$. We discretize the domain with $2^n$ points per variable. We denote the
$\ell^p$ norm of the function evaluated on $2^{Dn}$ grid points $\bm{x}$ with
\begin{equation}
    \Vert \bm{f} \Vert_p = \left( \sum_{\bm{x}} |f(\bm{x})|^p \right)^{1/p},
\end{equation}
and the $\ell^p$ norm filling fraction with
\begin{equation}\label{eq:filling_fraction}
    \mathcal{F}_p = \frac{\Vert \bm{f} \Vert_p}{2^{Dn/p} |f|_\text{max}},
\end{equation}
where
\begin{equation}\label{eq:f_max}
    |f|_\text{max} = \max_{\bm{x}\in [-1,1]^D} |f(\bm{x})|.
\end{equation}
Computing $|f|_\text{max}$ is hard in general, especially in higher dimensions.
Several prior works assume that this quantity is known or that a good upper bound $h \geq |f|_\text{max}$ can be found. Moreover, the cost of such preprocessing step is rarely discussed. An advantage of our protocol is that it does not require additional preprocessing steps. Let us assume that we want to implement a maximal degree-$d$ Fourier or Chebyshev
approximation $f_d$ to the target and we have computed coefficients $\bm{c}$.
Then we can use the following bound
\begin{equation}
    h = \Vert \bm{c} \Vert_1
    =
    \sum_{\bm{k}} |c_{\bm{k}}| \geq \max_{\bm{x}} | f_d(\bm{x}) |
    =
    |f_d|_\text{max}
\end{equation}
which only costs addition of $\mathcal{O}(d^D)$ numbers.

Inserting any bound $h$ in Eq.~\eqref{eq:filling_fraction} leads to a lower bound on the filling fraction
\begin{equation}
    \mathcal{F}_{p,h}
    =
    \frac{\Vert \bm{f} \Vert_p}{2^{Dn/p} h}
    \leq
    \mathcal{F}_p.
\end{equation}
Another useful quantity is obtained by replacing $f$ with the approximating function, such as a maximal degree-$d$
polynomial $f_d$ or a finite precision representation depending on the context. We denote with $\tilde{\mathcal{F}}_p$ an approximate $\ell^p$ norm filling
fraction.

\begin{table}
\footnotesize
\centering
\renewcommand{\arraystretch}{1.5}
\makebox[\textwidth][c]{
\begin{tabularx}{\linewidth}{c|>{\centering\arraybackslash}X>{\centering\arraybackslash}X>{\centering\arraybackslash}X>{\centering\arraybackslash}X}
    \toprule
    & Two-qubit gates
    & Success probability
    & Ancilla qubits
    & Function type
    \\ \midrule
    Black-box~\cite{groverSynthesisQuantumSuperpositions2000, Sanders_2019}
    & $\mathcal{O}(d^Dm^2)$
    & ${\cal F}_2^2$
    & $\mathcal{O}(d^Dm)$
    & any
    \\ \hline
    Grover-Rudolph~\cite{zalkaSimulatingQuantumSystems1998,grover2002creating, herbertNoQuantumSpeedup2021}
    & $\mathcal{O}(Dnd^Dm^2)$
    & 1
    & $\mathcal{O}(d^Dm)$
    & efficiently integrable distribution
    \\ \hline
    Adiabatic~\cite{Rattew_2022}
    & $\mathcal{O}(d^D m^2/\epsilon^2\mathcal{F}_1^4)$
    & $1-\mathcal{O}(\epsilon^2)$
    & $\mathcal{O}(d^D m + Dn)$
    & any
    \\ \hline
    MPS~\cite{schon_sequential_2005,iten_quantum_2016}
    & $\mathcal{O}(\poly(\chi) Dn)$
    & $1$
    & $\log \chi$
    & low entanglement
    \\ \hline
    FSL~\cite{Moosa2023}
    & $\mathcal{O}(d^D + Dn^2)$
    & 1
    & 0
    & Fourier series
    \\ \hline
    M-QSP~\cite{moriEfficientStatePreparation2024}
    & $\mathcal{O}(Dnd)$
    & $\mathcal{F}_2^2$
    & $1$
    & limited \& unclear
    \\ \hline
    LCU Fourier (this work)
    & $\mathcal{O}(d^D + Dn\log d)$
    & $\mathcal{F}_{2, \Vert \bm{c} \Vert_1}^2$
    & $D \lceil \log_2 (2d +1)\rceil$
    & Fourier series
    \\ \hline
    LCU Chebyshev (this work)
    & $\mathcal{O}(d^D + Ddn\log n)$
    & $\mathcal{F}_{2, \Vert \bm{c} \Vert_1}^2$
    & $D \lceil \log_2 (d+1)\rceil + D\lceil\log_2 n\rceil$
    & Chebyshev series
    \\ \bottomrule
\end{tabularx}
}
\caption{\label{tab:prev_work}
Summary of computational resources required to prepare a $Dn$-qubit quantum
state representing a $D$-dimensional target function.  For the oracles in the
black-box, Grover-Rudolph and adiabatic approaches we assume a maximal degree-$d$
polynomial approximation to the target stored in a register with $m$-bit
precision using the construction in Ref.~\cite{Haner2018}
(Eq.~\eqref{eq:oracle_complexity}).  $\chi$ is the bond order of the MPS
representation of the target function.  The success probability
$p_{\text{success}}$ can be amplified to 1 at the cost of multiplicative
overhead $\mathcal{O}(p_{\text{success}}^{-1/2})$ in gate count.
$\mathcal{F}_p$ is the $\ell_p$-norm filling fraction
(Eq.~\eqref{eq:filling_fraction}) and $\mathcal{F}_{2,\Vert\bm{c}\Vert_1} \leq
\mathcal{F}_2$. In practice, often only a bound on $\mathcal{F}_p$ is known,
which can lead to additional error. Note that the dependence of resources on
state preparation errors is mostly hidden in $d$, $m$ and $\chi$. For the
adiabatic algorithm, $\epsilon$ denotes the spectral distance to an exact state
preparation unitary.}
\end{table}

\subsection{Black-box algorithm}
\label{app:black-box}

In the \emph{black-box approach} we are given two oracles $O_\text{amp}$ and
$O_\text{rot}$. Oracle $O_\text{amp}$ writes a set of real numbers $\bm{\alpha} =
(\alpha_0, \alpha_1, \dots, \alpha_{2^n-1})$ with $0\leq \alpha_i \leq 1$ for
all $i=0, \dots, 2^n-1$ into an $m$-qubit register according to
\begin{equation}
    O_\text{amp} \ket{i} \ket{z} = \ket{i} \ket{z \oplus \alpha_i^{(m)}}.
\end{equation}
The first register has size $n$, $z$ is an $m$-bit integer and $\alpha_i^{(m)} =
\lfloor 2^m \alpha_i \rfloor$. Oracle $O_\text{rot}$ implements
\begin{equation}
    O_\text{rot} \ket{\alpha_i^{(m)}} \ket{0} = \ket{\alpha_i^{(m)}} \left( \sin\theta_i \ket{0} + \cos\theta_i \ket{1} \right)
\end{equation}
with $\theta_i = \arcsin(\alpha_i^{(m)}/2^m)$ so that $\sin\theta_i =
\alpha_i^{(m)}/2^m \approx \alpha_i$.  The goal is to prepare the state
$\frac{1}{\Vert \bm{\alpha} \Vert_2} \sum_{i=0}^{2^n-1} \alpha_i \ket{i}$. The
original black-box algorithm by
Grover~\cite{groverSynthesisQuantumSuperpositions2000} first performs
\begin{equation}
    (I^{\otimes n} \otimes O_\text{rot}) (O_\text{amp} \otimes I) (H^{\otimes n} \otimes I^{\otimes(m+1)}) \ket{0}^{\otimes n} \ket{0}^{\otimes m} \ket{0}
    = \frac{1}{\sqrt{2^n}} \sum_i \ket{i} \ket{\alpha_i^{(m)}}
        \left( \sin\theta_i \ket{0} + \cos\theta_i \ket{1} \right)
\end{equation}
and then uncomputes the second register by applying $O_\text{amp}$.
Post-selecting on the last register being 0 outputs the approximate target state
with success probability $p_\text{success} = \frac{\Vert \bm{\alpha}^{(m)}/2^m
\Vert_2^2}{2^n} \approx \frac{\Vert \bm{\alpha} \Vert_2^2}{2^n}$. It only
approximates the target state because of the $m$-bit precision of
$\alpha_i^{(m)}$ and the finite precision of computing the arcsine inside the
oracle (we assume the same precision $m$ for asymptotic scaling statements). The
original protocol prescribes amplitude amplification to boost the success
probability close to 1. Instead here we state the success probability of
post-selection without amplitude amplification to align with the presentation of
our algorithm. As noted in the main text, our algorithm -- or any probabilistic
algorithm discussed here -- could also use amplitude amplification at the cost
of increasing the number of gates by a factor of
$\mathcal{O}(p_\text{success}^{-1/2})$. The black-box protocol can be
generalized to complex amplitudes by considering two amplitude oracles, one for
the magnitude $|\alpha_i|$ and one for its phase~\cite{Sanders_2019}.

To connect with our work, consider $\alpha_i' = f_d(\bm{x}_i)$ with grid points
$\bm{x}_i \in [-1, 1]^D$ for $i=0, \dots, 2^{Dn}-1$ and $f_d$ a multivariate
polynomial approximation of $f$ with, without loss of generality, degree $d$ for
each variable. We use the rescaled amplitudes $\alpha_i = \alpha_i' /
|f_d|_\text{max}$ in the oracle $O_\text{amp}$ of the black-box approach. Then
the second register holds $f_d(\bm{x})/|f_d|_\text{max}$ with $m$-bit precision.
We assume that the $\mathcal{O}(d^D)$ coefficients of $f_d$ are computed in a
classical preprocessing step similar to ours. Then we get the following
resources for this oracle from Ref.~\cite{Haner2018}\footnote{For simplicity, we
only consider one polynomial over the entire domain instead of a piecewise
polynomial approximation. This does not affect asymptotic complexity.
Furthermore, it is conceivable that the quantum algorithm  for evaluating
polynomials in the monomial basis in Ref.~\cite{Haner2018} can be generalized to
evaluating polynomials in the Fourier and Chebyshev polynomials at similar cost
by using, \eg, the Fourier/Chebyshev polynomial recurrence relation or the
Clenshaw algorithm instead of a Horner scheme.}
\begin{equation}\label{eq:oracle_complexity}
    G_{\text{oracle}} = \mathcal{O}(d^Dm^2) \text{ two-qubit gates}
    \qquad \text{and} \qquad
    A_{\text{oracle}} = \mathcal{O}(d^Dm) \text{ ancillary qubits.}
\end{equation}
The same asymptotic complexity holds for computing the arcsine in $O_\text{rot}$
with $d$ replaced by the degree for a sufficiently accurate polynomial
approximation to the arcsine (and $D=1$). The oracle stores the computed value
in another register to desired precision (for simplicity assumed to be $m$).
Then its value is transduced to a phase by a series of controlled rotations,
which do not change the asymptotic complexity. Hence, the overall asymptotic
resources for both oracles are the same. The success probability is
\begin{equation}
    p_\text{success} = \frac{\Vert \bm{f}_d^{(m)} \Vert_2^2}{2^{Dn} |f_d|_\text{max}^2} = \tilde{\mathcal{F}}_2^2 .
\end{equation}
For sufficiently high precision $m$ and degree $d$ this success probability is
approximated by $\mathcal{F}_2^2$.

The preprocessing cost of the above implementation of the black-box approach is
the same as ours because both require computing coefficients for a maximal
degree-$d$ approximating polynomial. The black-box approach also requires
precomputing a bound on the maximum of $f_{d}$. This is an additional classical
preprocessing cost and the accuracy of this bound influences the required
precision $m$. As discussed at the beginning of Sec.~\ref{app:prior_works} we
could use the bound $\Vert \bm{c} \Vert_1$ using the coefficients of $f_d$. Then
the success probability of the black-box approach and this work (see
Table~\ref{tab:prev_work}) are essentially the same (up to e.g. differences due
to $m$-bit precision in the black-box approach).

The two-qubit gate count for the black-box approach is likely higher than ours.
This can be seen from the prefactors of the leading term $\mathcal{O}(d^D)$
($m^2$ vs 3 in our algorithm from the unitaries $A$, $A^\dagger$, $C$). The
number of Toffoli gates for several univariate functions in Ref.~\cite[Table
II]{Haner2018} suggests an order of magnitude of ten thousands for preparing
univariate functions. Gate counts for multivariate functions will be higher due
to the factor $\mathcal{O}(d^D)$. This should be compared to our explicit gate
counts, e.g. in
Figs.~\ref{fig:Ricker-chebyshev-results}-\ref{fig:Cauchy-fourier-results}. Note,
however, that this comparison is only indicative because we report optimized
\RZZ gate counts whereas Ref.~\cite{Haner2018} reports Toffolis. We also expect
the black-box approach to require considerably more ancillas than our approach
based on the explicit numbers in Ref.~\cite[Table II]{Haner2018} for univariate
functions and the favorable asymptotic scaling of our method in Table~\ref{tab:prev_work}.

\subsection{Grover-Rudolph algorithm}

The algorithm by Zalka~\cite{zalkaSimulatingQuantumSystems1998} applies to
target functions $f(x)=\sqrt{p(x)}$, where $p(x)$ is an efficiently integrable
probability density function. The algorithm was independently rediscovered by
Grover and Rudolph~\cite{grover2002creating} and is commonly called
Grover-Rudolph algorithm. A construction for efficiently integrable multivariate
distributions is given in Ref.~\cite{herbertNoQuantumSpeedup2021}.  At each step
$l=1, \dots, Dn$ of the algorithm an oracle encodes angles $\theta_i =
\arccos(\alpha_{l,i}^{(m)})$ for $i=0, \dots, 2^l$ into an ancilla register,
performs controlled rotations on a fresh qubit added to the main register, and
uncomputes the ancilla register. The angles are stored in an ancilla register
with sufficient precision (assumed to be $m$). The value of $\alpha_i^{(m)}$ is
determined by the integral over a part of the target distribution into an
ancilla register. The oracle is essentially $O_\text{amp}$ in the black-box
approach.

To compare resource requirements to our method we assume the same implementation
of $O_\text{amp}$ as in App.~\ref{app:black-box}. This does not take into
account resources for the coherent integration required for computing the
angles. The total number of two-qubit gates is higher by a factor $Dn$ than in
the black-box because of the iterative procedure of building up the main
register over $Dn$ steps. Hence, in general, we expect much higher gate counts
than in our algorithm (cf. the leading term in two-qubit gates in
Table~\ref{tab:prev_work}). The algorithm also requires complicated arithmetic
circuits to implement, while ours does not. We expect high numbers of ancillas
similar to the black-box approach, i.e. considerably higher than ours. The
success probability of Grover-Rudolph is 1, whereas ours is smaller in general.
The algorithm itself does not require classical preprocessing as the arithmetics
are absorbed in the oracle. However, an explicit oracle implementation may
require preprocessing.  For example, the implementation above requires finding
polynomial coefficients; likely with similar or worse preprocessing cost as
ours.  Variants of Grover-Rudolph have been proposed for the special case of
multivariate Gaussian distributions~\cite{kitaev2009wavefunction}; however,
explicit circuit constructions~\cite{bauer2021practical} showed that they
require more resources than generic state preparation methods
($\mathcal{O}(2^n)$ gates) in practice for moderate $n$.

\subsection{Adiabatic algorithm}

The adiabatic algorithm of Ref.~\cite{Rattew_2022} works for any complex
function $f$ assuming efficient pointwise evaluation.\footnote{The authors also
discuss an alternative state preparation for efficiently integrable functions
that could be beneficial in practice for discontinuous functions. Here we focus
on state preparation with pointwise evaluation.} It assumes access to the oracle
$O_\text{amp}$ defined in the black-box approach, App.~\ref{app:black-box}, to
evaluate and store intermediate function evaluations in an ancilla register with
$m$-bit precision.

The algorithm performs a discretized adiabatic evolution under the rank-1
Hamiltonian $H(s) \propto \dyad{\psi_{[s]}}{\psi_{[s]}}$ with $\ket{\psi_{[s]}}
\propto \sum_{\bm{x_i}}f_{[s]}^{(m)}(\bm{x}_i) \ket{\bm{x}_i}$ and the
parameterized function $f_{[s]}= (1-s) f_0 + s f$, where $f_0\propto 1$ has an
easy state preparation circuit, $f$ is target function of interest and
$0\geq s \geq 1$. Each step of the discretized evolution is implemented with a
sparse Hamiltonian simulation algorithm. The authors show that the total
evolution time is lower-bounded by a constant independent of $n$.  The number of
steps for error $\epsilon$ (in spectral distance to the exact unitary) is
$\mathcal{O}(1/\epsilon^2\tilde{\mathcal{F}}_1^4)$. The authors state that
knowledge of the filling fraction is not required to run
the algorithm. 

Resources are dominated by the oracle evaluating the function. For the
comparison in Table~\ref{tab:prev_work} we assume the implementation discussed
in App.~\ref{app:black-box}.  At intermediate steps $O_\text{amp}$ computes and
stores matrix elements
$A_{ij}(s)=(f_{[s]}^{(m)}(\bm{x}_i))^*f_{[s]}^{(m)}(\bm{x}_j)$ in an ancilla
register with $m$-bit precision. In addition, $\mathcal{O}(Dn)$ ancillas are
required to store the location of the nonzero entries of a 1-sparse matrix
required for the sparse Hamiltonian simulation. Compared to our algorithm,
adiabatic state preparation is nearly deterministic. Assuming the above oracle
implementation, the classical pre-processing cost is similar to ours. However, we
expect that an implementation of this algorithm requires considerably more
ancillas than our algorithm (see discussion in App.~\ref{app:black-box}).  We
also expect more two-qubit gates based on the larger prefactor of the term
$\mathcal{O}(d^D)$ ($m^2$ vs. 3 in our case).

\subsection{Matrix product states}

Matrix product states (MPS) algorithms are often heuristic in nature making a precise comparison
to non-heuristic quantum state preparation methods such as ours difficult. We
seek to represent a $D$-dimensional function discretized on $2^{Dn}$ grid points,
which requires an MPS with $Dn$ tensors.
Apart from the number of tensors, the complexity of an MPS is determined by its bond dimension $\chi$, and typical operations on the MPS will cost $\mathcal{O}(\poly(\chi)
Dn)$~\cite{orusPracticalIntroductionTensor2014}.
In the context of state preparation we
need to consider (i) the effectiveness and classical cost of representing a multivariate target function with an MPS, and (ii) the resources required to map the MPS to a quantum circuit.

We first discuss point (i). States representing smooth univariate functions with bounded derivative have
small von Neumann
entropy~\cite{garcia-ripollQuantuminspiredAlgorithmsMultivariate2021}. While
this suggests an MPS representation with small bond dimension, the equivalence is
not rigorous~\cite{schuch_entropy_2008}. On the other hand, univariate
degree-$d$ polynomials can be efficiently represented by an MPS with
$\chi=d+1$~\cite{grasedyck_polynomial_2010,oseledets_constructive_2013}. A way to extend those results to $D$-dimensional functions is to prescribe
the additional variables into the ordering of the tensors of an
MPS of size $Dn$. The ordering can
have significant influence on the bond dimension and it is unclear a-priori whether
a given multivariate function is well represented by an MPS. For example,
Ref.~\cite{garcia-ripollQuantuminspiredAlgorithmsMultivariate2021} observed a
increased, but relatively small, von Neumann entropy for 2D Gaussians. In contrast,
Ref.~\cite{lubaschMultigridRenormalization2018} observed a significant increase
of the von Neumann entropy with the dimension for the solutions of the nonlinear Schr{\"o}dinger equation, suggesting that the MPS is not well suited
for those functions.
Ref.~\cite{rodriguez-aldaveroChebyshevApproximationComposition2024} studied
low-rank approximations to a restricted class of multivariate functions that
avoids the exponential-in-$D$ cost $\mathcal{O}(d^D)$ of computing coefficients such
that they can be represented by MPS Chebyshev interpolants with low bond
order.  We note that similar low-rank approximations could also be used in our
quantum algorithm as a preprocessing step.

With regards to our point (ii) above, Ref.~\cite{schon_sequential_2005} constructs an exact MPS with $\chi$-level
qudits, which can be represented with qubits and a quantum circuit using
$\log\chi$ ancillas and depth $\mathcal{O}(\poly(\chi) Dn)$~\cite{iten_quantum_2016}.
Refs.~\cite{ran_encoding_2020,rudolph_decomposition_2022} propose heuristic
mappings from MPS to quantum circuits using $T$ layers of gates, for a total number of two-qubit gates scaling as
$\mathcal{O}(TDn)$, and without using ancilla qubits. 

There are very few end-to-end studies of quantum state preparation using MPS. Ref.~\cite{Holmes_2020}
considers smooth, real, univariate functions. They perform a piecewise polynomial approximation of the target function, translate
each degree-$d$ polynomial into an MPS, and add them up. The result is mapped to a quantum circuit using Ref.~\cite{ran_encoding_2020}. They assume $\chi=2$ approximates
the target function well enough on a $2^{n}$ grid which yields a two-qubit gate count of
$\mathcal{O}(n)$. The assumption $\chi=2$ is, however, in general not satisfied.
Ref.~\cite{Iaconis_2024} considers
univariate normal distributions and obtains comparable results. 

The discussion above suggests that, generally, the bond
order will increase in higher dimensions. Hence, in Table~\ref{tab:prev_work} we
state the more general two-qubit count $\mathcal{O}(\poly(\chi)Dn)$. Overall, we expect MPS to lead to shallower circuits than ours for smooth
univariate functions. For higher dimensions the comparison is less clear and
requires a dedicated study due to the heuristic nature of MPS algorithms. We emphasize that our multivariate state preparation method is not heuristic and it has a simple circuit construction.

\subsection{Fourier series loader}
\label{app:fsl}

The Fourier series loader (FSL)~\cite{Moosa2023} applies to complex multivariate
functions that can be written as a Fourier series. This is essentially the same
class of functions representable with our Fourier approach. FSL prepares a
truncated Fourier series of the form of Eq.~\ref{eq:f_d_Fourier_2d} (in arbitrary
dimensions $D$) on $2^{Dn}$ equidistant grid points on $[0, 1]^D$. A minor
difference in presentation is that Ref.~\cite{Moosa2023} also allows
discontinuous functions, whereas we chose to exclude them via our Lipshitz
assumption on $f$. The Fourier series of a discontinuous function exhibits large
errors (Gibbs phenomenon).  Ref.~\cite{Moosa2023} discusses heuristic
modifications of the Fourier coefficients to suppress those errors. We note that
our algorithm would also work if we loaded such modified coefficients at the
expense of the loss of fast uniform convergence guarantees and a small
preprocessing cost to compute them.

FSL first loads the $(2d+1)^D$ Fourier coefficients into a $(a=\lceil D
\log_2(2d+1) \rceil)$-qubit register. This is essentially a generic state
preparation circuit for arbitrary complex amplitudes similar in purpose to the
unitaries $A$, $C$, and $A^\dagger$  in our method. The authors suggest two suitable
implementations: uniformly controlled rotations~\cite{Mottonen_2004}, and
using the Schmidt decomposition of the coefficient state. Next, their algorithm prescribes a ladder of \CX on the remaining $Dn-a$
qubits controlled on the top qubit of the coefficient register. The application of one inverse quantum Fourier transform (QFT) per variable register of size $n$ yields the
desired state on the full register of size $Dn$. The $D$ inverse QFTs require
$\mathcal{O}(Dn^2)$ two-qubit gate, which is their leading gate count scaling in
$n$. We expect that their approach uses fewer two-qubits in regimes with small
$n$ and ours uses fewer in the high $n$ regime. Their algorithm is deterministic and uses no ancillas. On the other hand, our
protocols can also represent Chebyshev series.

\subsection{Multivariate quantum signal processing}
\label{app:mqsp}

The algorithm in Ref.~\cite{mcardle2022quantum} applies to univariate functions
$f\colon [a,b] \rightarrow \mathbb{C}$. Given a polynomial approximation of
$f((b-a)\arcsin(y) + a)/|f|_\text{max}$ their method implements this polynomial
with quantum signal processing (QSP).  Similar to our work, their algorithm works with continuous polynomial
approximations rather than storing a discretization in an auxiliary register,
such as in the black-box or Grover-Rudolph approach. Without amplitude
amplification the resource requirements are $\mathcal{O}(nd)$ two-qubit gates, $2$ (definite parity
polynomials) or $3$ ancillas (mixed parity) and $\tilde{\mathcal{F}}_2^2$ success probability. This success probability is
approximated by $\mathcal{F}_2^2$.

Ref.~\cite{moriEfficientStatePreparation2024} proposed a multivariate extension based
on a multivariate QSP (M-QSP)
ansatz~\cite{Rossi2022,mori2023comment,nemeth2023variants}. The implemented
function has the form $f(2\arccos(y_1)-1, 2\arccos(y_2)-1, \dots,
2\arccos(y_D)-1))$. The authors construct commuting block-encodings $U_j$ of
$\sum_{x_j} \cos(x_j) \dyad{x_j}{x_j}$ for each variable $j=1, \dots, D$, where
the sum runs over $2^n$ grid points of $x_j$. This can be done with one ancilla
and a total of $Dn$ controlled rotations.  The algorithm prescribes a length-$d$
sequence alternating between one of the block-encodings selected per step
followed by a single-qubit rotation on the ancilla plus one final single-qubit
rotation. The angles are determined by a maximal degree-$d$ polynomial
approximation of the target function. Post-selecting on the ancilla
being in state 0 gives the desired quantum state in the main register of size
$Dn$. The success probability is again $\tilde{\mathcal{F}}_2^2$.
Similar to the univariate QSP protocol in Ref.~\cite{mcardle2022quantum} the
multivariate protocol requires pre-computation of $|f|_\text{max}$ with the
caveats discussed above. 

While this proposal uses asymptotically fewer gates and ancillas than ours, the
class of functions representable by M-QSP is limited and largely not
characterized (except for the homogeneous, commuting bivariate
case~\cite{nemeth2023variants}). This is clear from a counting argument. A
$D$-dimensional maximal degree-$d$ polynomial has at most $\mathcal{O}(d^D)$
coefficients, whereas M-QSP only has $\mathcal{O}(d)$ free phases.  Moreover,
the classical cost and numerical stability of computing rotation angles for
general M-QSP is currently not well understood. In contrast, our algorithm works
for a wide class of multivariate functions and does not require this additional classical preprocessing.

\filbreak

\end{document}